\documentclass[11pt,preprint]{aastex}
\usepackage{longtable}
\usepackage{lscape}
\usepackage{color}
\pdfoutput=1
\everymath{\rm}
\everydisplay{\sf}

\newcommand{\tnm}[1]
   {\tablenotemark{#1}
   }

\slugcomment{Revision 2015-Feb-12: to be published in ApJ}
\shorttitle{Co-spatial UV-Optical Spectra of Galactic Planetary Nebulae}
\shortauthors{Dufour, Kwitter, Shaw, et al.}

\begin{document}

\title{Co-spatial Long-slit UV/Optical Spectra of Ten Galactic Planetary Nebulae with \textit{HST}/STIS 
I. Description of the Observations, Global Emission-line Measurements, and CNO Abundances\altaffilmark{1}}
\author{
Reginald J. Dufour\altaffilmark{2}, 
Karen B. Kwitter\altaffilmark{3}, 
Richard A. Shaw\altaffilmark{4}, 
Richard B.\ C.\ Henry\altaffilmark{5}, 
Bruce Balick\altaffilmark{6}, 
\and
Romano L.\ M. Corradi\altaffilmark{7,8}
}

\altaffiltext{1}{Based on observations with the NASA/ESA \textit{Hubble Space Telescope} obtained at the Space Telescope Science Institute, which is operated by the Association of Universities for Research in Astronomy, Incorporated, under NASA contract NAS5-26555.}

\altaffiltext{2}{Department of Space Physics and Astronomy, Rice University, Houston, TX  77251}
\altaffiltext{3}{Department of Astronomy, Williams College, Williamstown, MA  01267}
\altaffiltext{4}{National Optical Astronomy Observatory, Tucson, AZ  85719}
\altaffiltext{5}{Department of Physics and Astronomy, University of Oklahoma, Norman, OK 73019}
\altaffiltext{6}{Department of Astronomy, University of Washington, Seattle, WA 98195}
\altaffiltext{7}{Instituto de Astrof{\'{\i}}sica de Canarias, E-38200 La Laguna, Tenerife, Spain}
\altaffiltext{8}{Departamento de Astrof{\'{\i}}sica, Universidad de La Laguna, E-38206 La Laguna, Tenerife, Spain}

\begin{abstract}

We present observations and initial analysis from an \textit{HST} Cycle 19 program using STIS to obtain the first co-spatial, UV-optical spectra of ten Galactic planetary nebulae (PNe). 
Our primary objective was to measure the critical emission lines of carbon and nitrogen with unprecedented S/N and spatial resolution over the wavelength range 1150--10270 \AA, with the ultimate goal of quantifying the production of these elements in low- and intermediate-mass stars. 
Our sample was selected from PNe with a near-solar metallicity, but spanning a broad range in N/O based on published ground-based and \textit{IUE} spectra. 
This study, the first of a series, concentrates on the observations and emission-line measurements obtained by integrating along the entire spatial extent of the slit. 
We derived ionic and total elemental abundances for the seven PNe with the strongest UV line detections (IC~2165, IC~3568, NGC~2440, NGC~3242, NGC~5315, NGC~5882, and NGC~7662). 
We compare these new results with other recent studies of the nebulae, and discuss the relative merits of deriving the total elemental abundances of C, N, and O using ionization correction factors (ICFs) versus summed abundances. 
For the seven PNe with the best UV line detections, we conclude that summed abundances from direct diagnostics of ions with measurable UV lines gives the most accurate values for the total elemental abundances of C and N (although ICF abundances often produced good results for C). 
In some cases where significant discrepancies exist between our abundances and those from other studies, we show that the differences can often be attributed to their use of fluxes that are not co-spatial.
Finally, we examined C/O and N/O versus O/H and He/H in well-observed Galactic, LMC, and SMC PNe, and found that highly accurate abundances are essential for properly inferring elemental yields from their progenitor stars. 
Future papers will discuss photoionization modeling of our observations, both of the integrated spectra and spatial variations of the UV vs. optical lines along the STIS slit lengths, which are unique to our observations.

\end{abstract}


\section{Introduction} 

The elements carbon (C) and nitrogen (N) are present in all known life forms, and identifying the major stellar production sources of these elements is one of the most pressing problems in galactic abundance and astrobiology studies today. 
That C and N are synthesized and ejected by both massive stars, and by low and intermediate mass stars (0.8M$_{\odot}\le M \le8M_{\odot}$; LIMS), is not in doubt. 
The existence in the Galaxy of WC and WN stars with progenitor masses exceeding 20~M$_{\odot}$, as well as carbon stars and planetary nebulae (PNe) with LIMS as progenitors, suggests that the Galactic level of these elements is likely mediated by both components of the mass spectrum. 
The real challenge is to determine the proportional contribution that each component makes.

Numerous theorists have used stellar evolution models to predict the fraction of synthesized C and N, as a function of progenitor mass and metallicity, that LIMS eject into their PNe.
Historically, optical spectra of collisionally excited emission lines in PNe have been used to determine N and O abundances from the strong emission lines of [\ion{O}{2}]  $\lambda$3727 and [\ion{O}{3}] $\lambda\lambda$4959,5007 and [\ion{N}{2}] $\lambda\lambda$6548,6583. 
However in a majority of PNe N$^+$ represents only a very minor fraction of the total N abundance, and the ratio of O$^+$/O$^{++}$ has been used as an ionization correction factor (ICF) to estimate a value of N/O. 
In the higher ionization objects a correction is needed for {unobserved} O$^{+3}$, as indicated by the amount of He$^{++}$ compared to He$^+$, and the [\ion{N}{2}] lines are exceedingly weak, making the N abundance uncertain. 
For C, there are no collisionally excited lines in the optical spectral region at all. 
Only recently have investigators attempted to use weak recombination lines (RELs) of N, O, and C to derive CNO abundances, with most investigations finding significantly higher CNO abundances from RELs than 
from collisionially-excited lines (CELs) for O and N; this is the the well known ``abundance discrepancy factor" \citep[see, e.g., ][]{L06}.

However, in the ultraviolet there are strong collisionally-excited lines of the higher ionization states of N and several ionization states of C and O. 
Since the 1980Õs extensive observations of these lines in numerous PNe have been made with the \textit{International Ultraviolet Explorer} (\textit{IUE}) satellite [cf. the review by \citet{koeppen87}, for example] and provided the first empirical data on C/H in the shells ejected from LIMS, as well as more accurate N/H values. 
However, the \textit{IUE} had limitations on the accuracy for which the emission line could be measured, due to the vidicon camera detector having a limited 8-bit encoding capability and significant fixed-pattern noise. 
Moreover, in order to assure photometric accuracy, the UV observations had to be made with a large $10\arcsec \times 20\arcsec$ oval aperture which was essentially impossible to match to ground-based optical spectrometers. This mis-match of apertures affected all abundance calculations of PNe when UV- and optical-band spectra were analyzed in tandem.

Despite its limitations, numerous investigations of CNO abundances in PNe from \textit{IUE} observations were published during the 1980Õs and beyond. 
Early on \citet{dufour91} and \citet{perinotto91} compiled many of the earlier \textit{IUE} results for individual PNe to evaluate CN production and conversion in PNe of Peimbert Types I and II \citep{peimbert83}. 
\citet{kb94} used \textit{IUE} and optical data to measure abundances in a large sample of southern planetaries. 
\citet{rola94} compiled published UV and optical line strengths of carbon to determine the fraction of PNe which are C-rich. 
Kwitter and Henry combined reprocessed (NEWSIPS) \textit{IUE} spectra of 20 galactic PNe with their own optical data (supplemented by the literature) to determine abundances of C, N, and O \citep{HKH96,KH96,KH98,HKB00}. 
With the launch of the \textit{Hubble Space Telescope} in 1990, a new era of UV spectroscopy capabilities for nebular studies was born with the Faint Object Spectrograph (FOS), which featured linear detectors of high dynamic range for spectroscopy at both UV and optical wavelength regions through apertures of identical size. 
The replacement of FOS with the Space Telescope Imaging Spectrograph (STIS) in 1997 added a two dimensional longslit capability and combined UV-optical spectroscopy. 
This paper is the first to employ this new feature for a detailed study of CNO abundances in several Galactic planetary nebulae.

The goal of the project described in this paper is to measure accurate C, N, and O abundances in PNe using new \textit{HST} STIS observations spanning a wavelength range of 1150--10270~\AA. 
We observed 10 PNe representing a broad range in N abundance, but with overall metallicities close to solar. 
We present the details concerning the observations in Section~2. 
Section~3 contains the results regarding the abundances and nebular properties of each nebula, and we discuss the implications of these results in Section~4. 
Finally, our conclusions regarding our empirical results are presented in Section~5. 
In a subsequent paper (Dufour, et al. in preparation) we will compute photoionization models of each of our objects in order to derive the central star properties.
From these results we will derive the birth mass of each progenitor, combine it with our C and N abundances in the current paper, and evaluate several sets of published stellar model predictions of C and N abundances in PNe. 

\section{Observations}
\subsection{Target Selection}

The \textit{HST} Cycle 19 TAC awarded us 32 orbits to observe ten PNe with STIS. We strove for three objectives in establishing our target list: 
1) a narrow metallicity range (as measured by O/H) centered on the solar value; 
2) a large range in N/O; and 
3) the highest surface brightness (and good angular size when practical), 
all inferred from optical data employed in our earlier studies of Galactic PNe. 
We first identified a large set of potential STIS targets for their favorable observability (surface brightness, total flux through the slit, etc.). 
For science reasons noted above, we then selected PNe with roughly solar O/H abundances [8.55 $<$ 12+log(O/H) $<$ 8.80]. 
From these we selected semi-finalists with a wide range of N/O. 
Any very similar targets were culled using excitation, morphology, and electron temperature criteria in order to select 10 finalists requiring 32 STIS orbits.

The 10 finalists initially selected were: IC~418, IC~2165, IC~4593, NGC~2440, NGC~3242, NGC~5882, NGC~6537, NGC~6572, NGC~6778, and PB6 (ESO-213-7). 
In developing the ``Phase~I'' observation template, we identified permissible slit orientations, given the \textit{HST} observation windows, and chose ones near or on the central stars that covered a good mix of ionization structure (evident from narrow-band WFPC2 images in most cases) and high surface brightness rims, knots, etc. 
After submitting the observation template we received a ``red-flag" warning from the project scientist and STIS safety officer that we needed to provide evidence from \textit{IUE} UV spectra that each CSPN was ``safe'' for observation by the UV MAMA detectors with the low resolution UV gratings (G140L \& G230L) --otherwise we would have to move the slit centers 5$\arcsec$ or more away from the stars to protect the detectors. 
To put it mildly, this was a surprise and required an entire re-evaluation of our object selection and slit locations, based on STIS safety ``rules" versus optimum science input. 
After constructing \textit{IUE} spectra of our ten targets, we found that four had ``unsafe" central stars for which the STIS slit had to be at least 5$\arcsec$ away from the stars. 
These ``unsafe" PNe were IC~4593, NGC~3242, IC~418, \& NGC~6572. 
Given the large angular size of NGC~3242, we could place the slit 5$\arcsec$ from the central star without sacrificing much in the way of surface brightness and ionization structure, but the three other PNe had to be replaced by larger PNe with ``safe" central stars or large angular size AND similar N/O ratios as the original objects. 
After studying 
the Kwitter \& Henry PNe abundance database (hereafter, Òthe KH database,Ó comprising published results from observations over the last two decades by Kwitter, Henry and collaborators) 
and the \textit{IUE} spectral archives, we chose IC~3568, NGC~5315, and NGC~7662 as replacements.

Figure~\ref{Montage} shows the final ten PNe chosen with the slit positions and orientations that were possible given \textit{HST} observation windows and STIS safety constraints. 
The targets, coordinates and slit positions are given in Table~\ref{tab:Apertures}. 
Most images were obtained with \textit{HST}/WFPC2, in multiple passbands, and are presented in false-color. 
The images are all on the same spatial scale, with NE to the upper-left; see the figure caption for details. 
While in some cases we were able to include the central star in the slit, in others we were obligated to offset the slit by 5$\arcsec$.

Because of the length of time involved in obtaining approval for the UV observations on an individual object by object basis, and the changes in some of the PN targets requiring the development of a new Phase II template, the first observations were not begun until 2012 January (NGC 3242) and completed a year later (2013 January; NGC 2440). 

\subsection{Observing Strategy}

Our observing strategy was designed to achieve our primary science goals. 
The first goal was to obtain spectra of each target that covered the full, uninterrupted spectrum from UV through Optical (1150\AA\ through 10,150\AA), with sufficient resolution and sensitivity to derive nebular gas diagnostics and ionic abundances of all critical species. 
We achieved these goals by allocating a full orbit to obtaining UV spectra with the G140L, G230L, and G230M gratings, and two (or for the faintest targets, three) orbits to obtain optical spectra with the G430L, G430M, G750L, and G750M gratings. 
We balanced the need for good spectral resolution with the need for high sensitivity by using the 52$\arcsec$X0.2$\arcsec$ slit for the low-resolution gratings, and 52$\arcsec \times0.5\arcsec$\ for the medium-resolution gratings. 
The higher resolution gratings allowed us to resolve blends of a few critical lines: \ion{C}{3}] $\lambda1906$ from $\lambda1909$, \ion{H}{1} $\lambda4341$ from [\ion{O}{3}] $\lambda4363$; \ion{H}{1} $\lambda6563$ from [\ion{N}{2}] $\lambda6548$ and $\lambda6583$, [\ion{S}{3}] $\lambda6312$ from [\ion{O}{1}] $\lambda6300$, and [\ion{S}{2}] $\lambda6716$ from $\lambda6731$. 

The second goal was to make \textit{co-spatial} observations (i.e., with identical positioning and orientation of the apertures) of each target in all spectra, in order to avoid highly uncertain corrections for ionization stratification. 
The brightness limits (local and global) for the STIS/MAMA detectors, the required segregation of MAMA and CCD exposures to separate visits, as well as the need to provide some flexibility in the final slit orientation for scheduling reasons, presented some challenges to the design of our observing program. 
All of our targets are spatially extended (a few are larger than the spatial extent of the slit for MAMA exposures), most nebulae have high UV surface brightness, and many of the central stars are very bright in the UV. 
On the other hand, the bright, central portion of these nebulae is precisely where we expected to detect changes in ionization with position (a third goal of our observing strategy). 
We therefore constrained our MAMA visits to occur close in time to the initial CCD visits, in order to use the same guide stars for the MAMA visit aquisition.

\subsection{Data Acquisition}

The observations were executed between 2012-Jan-15 and 2013-Jan-28, with the UV (MAMA) and optical (CCD) visits occurring within a few days of each other. See the observing log for this program in Table~\ref{tab:ObsLog}. 
Most observations were successful, except for a failure with the NUV exposures for NGC~6537 (which was subsequently repeated some weeks later), and an acquisition failure for NGC~2440. 
All of the exposures for NGC~2440 were executed, but our examination of the acquisition images shows that the UV and Optical slit positions were not perfectly co-spatial. 
We analyzed, with the help of ST~ScI staff, the failed acquisition sequence for NGC~2440. 
The acquisitions begin with a short exposure of a 5\arcsec$\times5\arcsec$ field within STIS 50CCD aperture at the initial spacecraft pointing. 
The next step, locating the brightest feature in the field, failed for both the UV and CCD visits. 
The subsequent offset in each visit placed the final slit position for the UV and optical-band exposures in different locations. 
Figure~\ref{Acq_seq} shows the acquisition images at the initial pointing, and the final positions for the UV and Optical (northeast is in the upper-left). 
A green marker is superimposed on each image, indicating a nebular knot in common; the location of the knot indicates that the final location of the slit reference position in the optical-band lies almost directly south of the UV position by about 2.17\arcsec. 
In a forthcoming paper we will analyze the emission line profiles for the brighter nebulae in our sample. 
Our initial profile analysis for NGC~2440 suggests that the ionization stratification (and, perhaps, extinction) varies significantly with position, so that the ionic abundance analysis involving UV emission lines may not correspond well to that of the optical lines. 
While the emission line fluxes are presented here (see Sect. 2.4), it must be kept in mind that the UV and optical data are from different regions in the nebulae that may have somewhat different ionization.

\subsection{Data Reduction and Spectral Extractions}

The data were reduced and calibrated using the CALSTIS v2.36 pipeline (ca. 2011-May-27). 
The processing depends upon the detector in use (MAMA or CCD), and is described in detail by \citet{STIS_dhb}. 
Briefly, the processing includes overscan and bias correction (CCD only) or re-binning by a factor of 2 (MAMA only), bad pixel flagging, dark and flat-field corrections, fringe correction (CCD G750 only), calibrations of the world coordinate system (wavelength and spatial extent of the slit), flux calibration \citep[see][and references therein]{STIS_fluxCal}, and geometric rectification. 
The pipeline is capable of producing a variety of calibrated data products, but our spatially extended targets require custom spectral extraction from the calibrated, two-dimensional spectrograms: the \texttt{*\_sx2.fits} files (hereafter, SX2). 
These products include a science array, a bit-encoded mask array that records detector or processing anomalies at the pixel level, and a variance array. 

For each PN we examined the SX2 files and the emission profiles of the brighter lines from all gratings to determine the optimal spatial region for extraction. 
Since the spatial scales differ between the MAMA and CCD detectors (and between some CCD gratings), we developed a utility to extract regions that are spatially matched, rounded to the nearest whole pixel, for all gratings for a given target. 
The extraction regions are given in the last column of Table~\ref{tab:Apertures}, in arcseconds relative to the slit reference position. 
Note that the spatial extent of the extraction regions is rather large, spanning nearly the entire extent of the target, so rounding the extremes to integral pixels has no significant effect on the relative fluxes between gratings. 
The extractions to one dimension were derived from the SX2 images by averaging pixel values at each wavelength (column) over the specified spatial range, normalizing by the extraction area in pixels, and converting from surface brightness to flux density. 
All pixels marked as \textit{bad} in the pixel mask were excluded except for those flagged as affected by flat-field blemishes, noisy background, and excessive dark rate, as long as no other flags applied. 
While ignoring these flags adds some additional uncertainty to the average value, the final average is usually dominated by well exposed pixels where these pathologies are unimportant. 
In rare instances (at the edge of the MAMA detectors) all pixels were flagged as bad, so the average flux at such wavelengths was set to zero. 
No background subtraction was performed at this stage. 
In cases where the slit was positioned close to or included the central star, special care was taken to minimize stellar contamination of the nebular spectrum.
In the end, signal-to-noise considerations and scattering of the starlight by dust in some nebulae compelled us to accept some stellar light in the cases of IC~3568, NGC~5315, and NGC~5882.   

\subsection{Measurements}

Emission-line fluxes were measured using the {\it IRAF}\footnote{{\it IRAF} is distributed by the National Optical Astronomy Observatory, which is operated by the Association of Universities for Research in Astronomy (AURA) under cooperative agreement with the National Science Foundation.} \textbf{splot} package. Most lines were observed only in the {\it L} gratings. 
Regions with important closely-spaced lines were also observed with the medium-resolution {\it M} gratings, through a wider slit: see Table~\ref{tab:ObsLog}. 
Measurements of the better-resolved {\it M} spectra yielded ratios for these lines, which were then used to apportion the summed flux measured on the {\it L} spectra. 
For weak lines detected only in the {\it M} spectra, the measured flux was multiplied by 2.5, the ratio of the {\it L} and {\it M} slit widths. 
Figures~\ref{spectrum1} and \ref{spectrum2} show UV and optical spactrograms, respectively, for IC~2165 as examples of the data quality and spectral range. Numerous emission lines are identified; inset graphs show enlargements of some line complexes. 

Lines intensities are given in Tables~\ref{tab:Flux1} and \ref{tab:Flux2}, normalized to F(H$\beta$)=100. 
The columns are, in order: the wavelength of the emission line; the line identification by ion; \textit{f}($\lambda)$, the value of the reddening function at that wavelength; F($\lambda$), the observed line flux; and I($\lambda$), the corrected line intensity with its uncertainty. 
Stellar and stellar+nebular composite lines are noted, as are lines that may be affected by artifacts in the 2D spectrogram. 
At the end of the table are, for each nebula, the calculated value of \textit{c}, the logarithmic reddening parameter; the expected, zero reddening ratio of F(H$\alpha$)/F(H$\beta$) for the derived nebular temperature (from T[\ion{O}{3}]) and density (usually from N[\ion{S}{2}]); and the observed F(H$\beta$) through the extraction window. 
Note that we give the line intensities for NGC~2440 in two columns: one for the UV (MAMA) spectra and another for the optical (CCD) spectra. 
Although the regions are in principle distinct, we normalize the fluxes to F(H$\beta$) as measured in the optical, and determine the diagnostics and abundances as if they were from the same location (see \S3.2.3); we call out the potential for problems in the results below.  

Uncertainties were estimated in the following way. 
For each line, we obtained the RMS continuum value on either side (the \textit{m} key in \textbf{splot}); sometimes only one side was amenable for measurement. 
We took the average RMS of the two sides and multiplied it by the line FWHM (\textit{k} key or via deblending option in {\it splot}) to obtain the line flux uncertainty. 
The emission-line fluxes along with their uncertainties constitute the input for our abundance determinations using ELSA, our 5-level atom code \citep{J06}. 
ELSA propagates the uncertainties though the calculations, including the intensities, diagnostics and abundances. 
The first step in the analysis is to generate a table of line intensities that have been corrected for interstellar reddening and for contamination of the hydrogen Balmer lines by coincident recombination lines of He$^{++}$. 
In addition, ELSA can disentangle some unresolved line blends (here, [\ion{Ne}{3}] $\lambda$3968 and H$\epsilon$ $\lambda$3970) when one or both of the lines has a known ratio to another measured line. 
We corrected for the effects of reddening using the function of \citet{savage79} in the optical region, and \citet{seaton79} in the ultraviolet. 
Details of the analysis using ELSA  are described in \citet{Milingo10}.

A few very weak emission lines were noted in three objects that we were unable to identify. 
These are noted in Table~\ref{tab:ULines} by wavelength and the nebula in which they were found; the intensities were comparable to the noise in the surrounding continuum. 
Though weak, the lines appear in the SX2 images to have profiles similar to other, well exposed nebular lines rather than artifacts (e.g., hot pixels or charge trails). 
We note them here in the hope that future investigations may be able to make use of them. 

\section{Results}
\subsection{Plasma Diagnostics and Abundances}
 
We present the plasma diagnostics in Table~\ref{tab:Diags}. 
The [\ion{O}{3}] temperature is derived from the I($\lambda4343$)/I($\lambda5007$) ratio. 
Where available, the [\ion{N}{2}] temperature is derived from the I($\lambda5755$)/I($\lambda6584$) ratio; otherwise, based on previous work \citep{KH01}: if \ion{He}{2} $\lambda4686$ is detected, as it is in all PNe here, we adopt the carefully derived result from \citet{kaler86} that applies under this condition, i.e., T[\ion{N}{2}] = 10,300~K. 
If the required lines have been detected, we also report values of T[\ion{O}{2}] from I($\lambda7323$)/I($\lambda3727$), 
but owing to the high uncertainty associated with these derived temperatures they are not used in any calculations. 
The T[\ion{S}{2}] diagnostic from I($\lambda[4068+4076]$)/I($\lambda[6717+6731]$) failed to give reliable results because the emission is quite weak, and the $\lambda4068+4076$ doublet is potentially blended with weak recombination lines. 
If the [\ion{S}{3}] lines $\lambda6312$ and one of $\lambda9069$ or $\lambda9532$ are available, T[\ion{S}{3}] is calculated and, if it is within  5000~K of T[\ion{O}{3}], we use it  to derive the abundances of S$^{+2}$ and Cl$^{+2}$. 
In general, T[\ion{O}{3}] is used for both helium ions and for other ions in states +2 or above; T[\ion{N}{2}] is used for the other singly-ionized species. 
Electron densities are calculated using ratios of [\ion{S}{2}] I($\lambda6717$)/I($\lambda6731$) and \ion{C}{3}] I($\lambda1909$)/I($\lambda1906$). 
If only one density diagnostic is available, it is used for all calculations. 
If both are available, N[\ion{S}{2}] is used to calculate abundances of singly-ionized species, and N[\ion{C}{3}] is used for the higher-ionization species. 

Ionic abundances derived using ELSA are given in Table~\ref{tab:IonicAbund}. 
The first column lists the ion and wavelength used to calculate the values in each row; the adopted value for the ionic abundance corresponds to the mean of all the observed lines of that ion, weighted by raw observed flux, and is used in subsequent calculations. 
Measured lines contaminated by stellar emission (indicated in Tables~\ref{tab:Flux1} and \ref{tab:Flux2}) are excluded from further analysis. 
Values of the ionization correction factor (ICF) that was derived to compute total abundances are shown at the end of each ion listing; these have been calculated in ELSA as described in \citet{KH01}, except for carbon, which was not studied in that paper. 
Here we use the following: 

$$C/H = \frac{C^{++}}{H^+} \times ICF(C)$$ where
$$ICF(C) = \frac{(O^{+} + O^{++})}{O^{++}} \times \frac{(He^{+} + He^{++})}{He^{+}}$$

The total elemental abundances are shown in Table~\ref{tab:ElemAbund}. For C- and N-related parameters, we show both the values derived using ICFs and those obtained by summing abundances of observed ions. 
Note that abundances of O, Ne, S, and Ar presented here are ICF values. 
The last two columns give values for the Sun \citep{A09} and Orion \citep{esteban04}.

\subsection{Individual Nebulae}

Here we discuss the abundance results for eight of the 10 PNe we observed. 
NGC~6537 and NGC~6778 are not included since the low S/N of the STIS data prevented any meaningful homogeneous abundance analysis. 
The emission line intensities for these excluded objects, presented in Table~\ref{tab:Flux2}, should inform future investigators of the stronger UV emission lines, and of the approximate scaling from the UV to optical band. 
Our spectrograms for PB~6 are also weakly exposed, but we are able to combine the UV emission lines with published spectra to confirm the very high enrichments of He, C, and N noted in the literature.
Photoionization models of these eight PNe will be presented in a forthcoming paper. 

We note again that our data are strictly co-spatial (except for NGC~2440), meaning that we are sampling the same region in each PN across the entire observed spectral range. 
The co-spatial results are thus free of any need for aperture size or placement corrections whose values can be difficult to calculate and whose effects on the final abundances almost impossible to assess. 
All calculated elemental abundances are listed in Table~\ref{tab:ElemAbund}; we only discuss CNO abundances here. 
Table~\ref{tab:AbundComp} shows our CNO abundances and those of various other authors with whom we compare below. 
All of our O abundances come from ICF calculations. 
We compare our summed values for C and N where available (for some PNe we had to exclude lines compromised by stellar emission, which we discuss in more detail below). 
We derived upper limits for key undetected ions and note their potential contribution. 
We also compare the abundances derived with the ICF values, and comment on any discrepancies with the sums. 
Note that the ICF values for PNe where the value of T[\ion{N}{2}] is assumed are particularly uncertain; however, our conclusions are based on the summed-ion results, which are unaffected by this uncertainty. 
As mentioned above, T[\ion{N}{2}] is used only to calculate ionic abundances for singly-ionized species, which are generally minor contributors to the total C and N abundances. 
Further, comparison of results using our default T[\ion{N}{2}] of 10,300~K from \citet{kaler86} with the recipe from \citet{kb94} yields insignificant differences in the contribution of N+ to the total N abundance. 

In the following subsections where we compare our CNO abundances for each nebula in detail with those from other authors, it is worth noting at the outset the level of agreement among all authors for all nebulae, shown in Fig.~\ref{AbundComp}. 
Individual research groups are denoted with the same symbol, as they tend to use the same methodology, the same atomic parameters, and similar observing technique. 
The elements are differentiated by symbol color. 
The figure shows that the agreement among all authors for O abundance is generally good, within 30\% for most cases. 
The agreement for C and N is much poorer, with many deviations approaching a factor of a few. 
In an exhaustive comparison of abundances derived for Magellanic Cloud PNe, \citet{Shaw10} noted several reasons why abundances of the same object often differ from author to author. 
In some cases the differences in Fig.~\ref{AbundComp} can be attributed to other authors' use of fluxes obtained with different apertures; for nebulae with significant ionization stratification such discrepancies are worrisome. 
Discrepancies can also arise from the use of different atomic parameters, different extinction constants, different techniques (e.g., the adoption of different T$_e$ or N$_e$ for different ions), or flawed observing techniques or data calibration. 
Note that discrepancies in any datum or derived parameter propagates downstream to the derivation of the final elemental abundances. 
This is a particular problem for abundances derived with ICF methods. 
In the comparisons below, we attempted to select data from the literature that represent the best available for these well-observed nebulae. 

\subsubsection{IC~2165}
This object has a bright, angularly small ($\sim7$\arcsec) core and a faint, extended halo \citep{Corradi03}. 
There is significant ionization stratification with position along the slit, with most of the emission from very low ionization species such as O$^+$, N$^+$, and S$^+$ located on the periphery (see Fig.~\ref{Montage}). 
Our ICF and summed ion abundances agree very well for both C (12\% difference) and N (7\% difference), indicating that the ICF method is doing a good job of accounting for ionization states that would be missed had only optical data been available. 
\citet{KHM03} observed IC~2165 (optical only); their O/H value is within $\sim$$12\%$ of ours. 
Their N/H value is $\sim$60\% larger than our N/H, due to their comparably greater ICF(N) stemming from a smaller derived O$^+/H^+$. 
They did not observe C. 

\citet{PBS04} reported abundances for IC~2165 derived from combining \textit{IUE}, ground-based optical (primarily) from \citet{Hyung94}, and \textit{ISO} fluxes. 
Both the \textit{IUE} and \textit{ISO} apertures are larger than this nebula, but the ground-based data were averages from multiple observers, where some spectra were obtained through smaller apertures placed at different nebular locations. 
The optical slit spectra from \citet{Hyung94}, which was only 4\arcsec\ long, weighted the center of the nebular emission more heavily and hence missed much of the [\ion{O}{2}] $\lambda3727$ and [\ion{N}{2}] $\lambda6583$ emission. 
Their O abundance is very close to ours, but this is somewhat of a coincidence: they derived a much lower O$^+$ abundance because of a lower observed I(3727), but they derive a significant O$^{+3}$ abundance from the IR $25.8~\mu$m line. 
We observed \ion{O}{4}] UV emission, but we used ICF(O) to account for O$^{+3}$ and higher ionization stages. 
Their N/H is $\sim$17\% less than ours, mostly because their I(6584) was much lower than ours. 
Finally, their C/H is $\sim$50\% more than our value, in spite of very similar intensities for the relevant emission lines: the discrepancy may come from either or both of different atomic parameters or their adoption of somewhat different T$_e$ values. 

Most recently, IC~2165 was observed by \citet{BRD13} (optical only), who obtained optical echelle spectroscopy of the inner 22\arcsec$\times27$\arcsec\ region. 
The much narrower STIS slit, as shown in Fig.~\ref{Montage}, passes through the long axis of the nebula, and does not sample as large a variety of environments. 
Comparing with their results derived by the ``standard" method, our O abundance is 37\% larger than theirs, due primarily to our higher O$^+$/H$^+$. 
Our value and their value for N$^+$/H$^+$ are within $\sim$$12\%$, but as a result of their larger ICF(N), our N/H abundance is only $\sim$55\% of theirs. 
Their C abundance, derived exclusively from permitted lines, is almost double ours. 
The He abundance (He/H=0.106), coupled with low N/O ($\sim$0.3), corroborates the conclusion that IC~2165 originates from a progenitor not much more massive than $\sim$2M${_\odot}$ (e.g., \citealt{BRD13}). 
The C/O ratio ($\sim1.2$) is larger than the solar value, perhaps indicative of some C production.

\subsubsection{IC~3568}
This object is $\sim18$\arcsec\ in size, with a bright inner core of $\sim7$\arcsec; there is little ionization stratification, however the FUV spectrum shows stellar P-Cyg profiles in \ion{N}{5} and \ion{C}{4}. 
Our calculations of ICF and summed-ion abundances for C agree to within 6\%, giving confidence in the numbers. 
Many of our spectrograms for this object were less than optimally exposed: we determined only upper limits for nebular UV N lines, for example. 
\citet{HKB04} derived abundances for IC~3568 from optical spectra; their N/H and O/H are 37\% and 28\% higher, respectively. 
\citet{L04a, L04b} also observed IC~3568 and calculated CNO abundances combining \textit{IUE} and \textit{ISO} data with optical recombination lines and with collisionally-excited lines. 
We compare our results with their collisionally-excited line results and find that the C/H and O/H agree within 25\%. 
However, our N/H using ICF(N) is only 40\% of theirs, despite our N$^+$/H$^+$ being twice as large. 
The difference is due to their inclusion of \ion{N}{3}] $\lambda$1750 from \textit{IUE} data (we have only upper limits from our STIS data) to derive N$^{++}$/H$^+$. 
We detected only the N$^+$ ion which had a rather large ICF(N) of 33.3, but our sum of N$^+$ and the upper limit to N$^{+2}$ agrees well with that of \citet{L04a, L04b}. 
Our O abundance agrees within $\sim$25\% to that of \citet{HKB04}. 
Based on our results, He/H (0.118) is slightly above solar, C/O ($\sim$0.5) is solar, and N/O (0.04) is slightly sub-solar. 

\subsubsection{NGC~2440}
This object is angularly large ($74\arcsec \times42\arcsec$) with highly stratified ionization. 
Our results for NGC~2440 are uncertain to the extent that the slit positions for the UV and optical observations sampled different locations (see \S~2.3). 
Based on the STIS data, the ICF abundances of C and N are 60\% higher and 2.3 times higher than the ion-sum values, respectively. 
Both O$^+$ and O$^{++}$ are well measured in our spectra, as are He$^+$ and He$^{++}$, so the ICF values should be accurate. 
It may be that the spatial offset of the UV and optical spectra, described above, sampled sufficiently different regions that these two results are not directly comparable. 

NGC~2440 has been observed by many authors, including \citet{BS02} who combined \textit{ISO} IR data with \textit{IUE} UV data and previously-published optical data; and \citet{Tsamis03}, who observed the entire nebula and combined their optical data with \textit{IUE} data; \citet{KHM03} and \citet{Krabbe06} observed in the optical only. 
Our ion-summed C abundance is 82\% of what \citet{Tsamis03} reports, and only 28\% of values reported by \citet{KHM03} and \citet{BS02}. 
Our ion-summed N abundance is $\sim$50\% of \citet{KHM03}'s, $\sim$68\% of \citet{Krabbe06}'s, and $\sim$95\% of \citet{BS02}'s, but 2.8 times that of \citet{Tsamis03}. 
These last authors did not detect any N$^{+}$, so their entire N abundance is based on \textit{IUE} lines. 
Inspection of their ionic abundances reveals that their values of N$^{++}$, N$^{+3}$, and N$^{+4}$ are systematically 2--5 times lower than ours. 
The source of the discrepancy may lie in our mis-matched apertures. 
However, since this nebula is larger than the \textit{ISO} and \textit{IUE} apertures and is highly stratified, the discrepancy could equally well indicate a problem with matching the satellite data to those from the ground-based spectra. 
The He/H ratio in NGC~2440 (0.127) and the N/O ratio, whether ICF (1.6) or summed-ion (1.0) are both greater than solar and suggest that the progenitor of this PN was relatively high mass. 

\subsubsection{NGC~3242}
This object is angularly large, with a core of $16\arcsec \times 26\arcsec$, surrounded by a faint shell of about $40\arcsec \times35\arcsec$ \citep{Corradi03}. Within the bright inner core there is moderate ionization stratification. 
Our ICF and summed-ion results for NGC~3242 are quite disparate: C(ICF) $=1.5 \times$C(sum) and N(ICF) $= 0.5 \times$N(sum). \citet{MHK02} observed NGC~3242 in the optical only, as did \citet{Krabbe06}. 
\citet{PBS08} combined \textit{ISO} IR data with \textit{IUE} UV data and previously-published ground-based optical data. 
\citet{Tsamis03} observed the entire nebula and combined their data with \textit{IUE} data. 

All of these authors' O abundances agree with ours to $\sim30$\%. 
The C abundance from \citet{Tsamis03} is half of our ICF value and 73\% of our summed-ion value; they used only \ion{C}{3}] $\lambda$1909, and did not include \ion{C}{4} $\lambda$1549 or \ion{C}{2}] $\lambda$2324. 
The C abundance (and also the C$^{+2}$ and C$^{+3}$ ionic abundances) from \citet{PBS08} agrees perfectly with our summed-ion value \textit{in spite of significantly different intensities for the lines used to derive the ionic abundances}. 
Their C abundance is 70\% of our ICF value. 
Our ICF-based N abundance agrees to $\sim$20\% with those of \citet{Krabbe06}, \citet{Tsamis03}, and \citet{MHK02}, but is three times lower than that of \citet{PBS08}. 
Since our summed-ion N value is twice as high as the ICF value, the agreement just described suffers by a factor of two. 
The He/H ratio in NGC~3242 is slightly above solar (0.115), the N/O is 0.1 (ICF) or 0.2 (sum). 

\subsubsection{NGC~5315}
This object is angularly small ($\sim15$\arcsec), with modest ionization stratification, but there are very bright stellar emission lines. 
This analysis for this PN is problematic: the S/N is not as high as needed for a robust abundance analysis, since we had to avoid much of the spatial extent of the nebula in order to exclude a substantial contribution of scattered stellar P-Cyg emission to the nebular emission. 
As a result, we have not used any \ion{He}{2} lines in the abundance analysis; the ICF values in Table~\ref{tab:IonicAbund} should therefore be viewed as lower limits.
NGC~5315 was observed previously in the optical by \citet{Milingo10} and by \citet{Tsamis03} who observed the entire nebula, and combined their data with \textit{IUE} data; \citet{P02} combined their IR observations with \textit{IUE} and previously-published optical data. 
These other investigators derived O abundances $\sim$18-70\% larger than ours. Their N abundances are within 30\% of our N(ICF) values and 2-3 times our summed-ion N abundance. 
Our ICF and summed-ion C abundances are in excellent agreement, and within $\sim$10\% of the \citet{Tsamis03} value. 
Despite \citet{P02} deriving ionic abundances for C$^+$ and C$^{++}$ similar to ours, their application of an ICF yields a C abundance roughly twice ours. 
\citet{Milingo10}, using only the C$^{++}$ permitted line at $\lambda$4267, derive a C abundances 3.5 times ours.
The He/H ratio (0.132) and the N/O ratio (1.1, using the ICF value are clearly greater than the solar values.

\subsubsection{NGC~5882}
This nebula is intermediate in size \citep{corradi00} with a bright, elliptical inner shell ($11\arcsec \times6\arcsec$), surrounded by a spherical outer shell (15\arcsec); it has substantial ionization stratification. 
Scattered stellar light was an issue in the STIS observations, reducing the region of the slit usable for nebular analysis, with the higher-ionization lines of N and C being of stellar origin. 
NGC~5882 was observed in the optical by \citet{KHM03} and by \citet{Tsamis03}, who observed the entire nebula, and combined their data with \textit{IUE} data. 
\citet{PBS04} observed the full spatial extent of the nebula with \textit{ISO} in the IR, and combined their observations with \textit{IUE}  data in the UV and previously-published, ground-based optical data (primarily from \citet{Tsamis03}). 
All of the O abundances agree with ours to 25\%. 
The N abundances from these authors agree with each other to 20\% but are 2--3 times ours, which is based solely on optical N$^+$/H$^+$ and an ICF(N) of $\sim$33. 
Our upper limit to the \ion{N}{3}] $\lambda1750$ emission is less than half that of \citet{PBS04}, but their N$^{+2}$ abundance (the dominant ionization stage), derived from the $57\mu$m emission line, is larger than ours by a factor of 5. 

\citet{Tsamis03} and \citet{PBS04} derived C abundances from \ion{C}{3}] $\lambda$1909 in the same \textit{IUE} data, calculating similar C$^{++}$/H$^+$ ratios, about twice what we derive from the co-spatial STIS data. 
This, together with the use of different ICF values, produces total C abundance that are 75\% larger and 2.5 times larger than ours, respectively. 
The He/H (0.11) and N/O (0.13) in NGC~5882 is roughly solar

\subsubsection{NGC~7662}
This nebula is moderately large, with a bright shell $\sim30$\arcsec\ in diameter \citep{Corradi03}, and considerable ionization stratification. 
Our STIS data show that the ICF values and summed-ion values agree very well, within $\sim$$1\%$ for C and $\sim$$14\%$ for N, indicating that, as in IC~2165, the ICF method is yielding reliable abundances. 
NGC~7662 has been observed by \citet{KHM03} in the optical, and by \citet{L04a,L04b} who combined their data with \textit{IUE} UV data. 
The O abundances agree well, within 20\%. 
Our C abundance is only half that found by \citet{L04a,L04b}, despite our ionic abundances agreeing with theirs to within 30\%. 
The difference is that they apply an ICF (=2.0) to their ion sum. 
Our N abundance is $\sim$65\% of those derived by \citet{L04a,L04b} and \citet{KHM03}. 
The difference with \citet{L04a,L04b} appears mainly to be their higher N$^{+4}$ because of a higher observed \ion{N}{5} $\lambda1240$ emission, which is likely the result of scaling the fluxes from the \textit{IUE} large aperture to their 1\arcsec\ optical slit. 
Our ICF N abundance hinges on weak O$^+$/H$^+$, but the good agreement with the ion sum implies that the ICF itself is not the reason. 
The He/H (0.122) is higher than solar but the N/O ratio (0.13) very close to solar, 

\subsubsection{PB~6}

This object is a small ($14.0\arcsec \times 12.5\arcsec$), clumpy nebula with a bright, [WC] central star, as shown in Fig.~\ref{fig:pb6}. 
A ground-based image in [\ion{O}{3}] obtained by one of us (R.L.M.C.; included in Fig.~\ref{fig:pb6}) shows what appear to be two concentric, nearly circular shells, but the higher-resolution STIS image resolves the structure into a complex of knots. 
Some of the knots appear to have a linear, or cometary structure oriented away from the central star, not unlike those seen in Abell 30 and Abell 78 \citep{Fang_etal14}. 
The He/H, C/O, and N/O ratios for this nebula have been reported to be extraordinarily high. 
\citet{kaler91} first noted the presence of very strong \ion{O}{6} emission in the optical, and strong stellar emission is apparent in our UV and optical spectrograms as well. 

Our spectrograms are too weakly exposed to derive robust abundances, or even reliable physical diagnostics. \citet[][hereafter GRPP]{grpp09} obtained deep optical spectra in order to use faint recombination lines to derive abundances. 
They derived values for T$_e$ and N$_e$ from CELs which are consistent with other published values, and found from ORLs high enrichments of He, C, and N. 
Our emission line intensities, uncertain as they are, are consistent with those reported in GRPP apart from our detection of a stronger [\ion{N}{2}] $\lambda6583$ flux. 
Our spectrograms, and the two zones within PB~6 analyzed by GRPP, indicate that the ionization is not strongly stratified within this nebula. 
We used the optical intensities from GRPP and our UV data to compute the abundances from CELs using ELSA, in order to compare techniques (in spite of the mis-matched apertures). 
We find that the He and CNO abundances derived in this way generally agree with those from GRPP, although the strong observed (but uncertain) C$^{+3}$ abundance from our UV spectrum suggests that C could be even higher. 
\citet{keller14} modelled the central star spectrum to derive T$_{eff} = 165,000$~K and $log(L/L\sun) = 3.43$, using data from \textit{FUSE} and our UV spectrograms from STIS; their results are also consistent with the high ionization of this  nebula.

\section{Discussion}

It is well known that PNe exhibit a wide variety of ionization structures. 
This is especially true of PNe that are not fully ionized, where the ionization is often a complicated function of position within the nebula. 
Unless spectra can be obtained that intercept the light from the full spatial extent of the nebula (which is usually achieved only for angularly small PNe, such as those in external galaxies), the derivation of nebular conditions and elemental abundances must of necessity sample a limited, possibly small volume within the nebulosity. 
We have shown here (as many before us have) that the most reliable elemental abundances are derived by observing a full range of ionic species, which requires spectra spanning at least the UV through optical bands, at high S/N and moderate-to-high spectral resolution. 
Often this requires the use of multiple spectrographs on different platforms. 
We believe the most accurate abundances are derived when all spectroscopic apertures sample identical three-dimensional volumes of the nebulosity. 
Indeed, for nebulae with stratified ionization, co-spatial apertures are essential for deriving consistent plasma diagnostics, ICFs, and elemental abundances. 
We have shown in some key cases (including IC~2165, NGC~3242, and NGC5882) that the primary reason for the large difference between our derived ionic and elemental abundances, and those in the literature (up to a factor of a few: see Fig.~\ref{AbundComp}), is the mis-application of data from spatially distinct regions. 
It is our assertion, to be explored further in forthcoming papers, that to reach the level of precision necessary to discriminate between modern, competing theories of elemental yields in post-AGB stars, it simply will not suffice to scale or select emission line intensities from one nebular region and to transplant them to another, spatially distinct region. 

Figure~\ref{fig:4plex} graphically displays the summed abundance results given in Table~\ref{tab:ElemAbund} for the program PNe discussed in the previous section (large filled circles). 
Because our optical line measurements for PB6 proved to be unreliable (see \S3.2), the positions for PB~6 in all four panels were determined using abundances from \citet{pena98}. 
Despite the fact that our data for PB6 are less than satisfactory, we decided to include it in Fig.~\ref{fig:4plex} because it harbors a [WC] central star, it is a Type~I PN, and possesses an inordinately high nebular He abundance, making it an interesting object to compare with the other sample members.
The left-hand panels show log(C/O) and log(N/O) versus He/H, while the right-hand panels do so for the same dependent variables against log(O/H)+12. 
To provide a comparison with a larger sample, the small, gray dots show abundance ratios from {the KH database. 
In the right-hand panels we have also added results from \citet{stanghellini05,stanghellini09} and \citet{stasinska98} for the LMC and SMC in order to extend the metallicity range to lower values. 
Within our sample of eight PNe, Type~I objects are identified with italicized font, while names of non-Type~I PNe are shown with regular font. 
The bold dashed lines show solar values from \citet{A09}. 

The right-hand panels demonstrate that there is no clear trend in either C/O or N/O with respect to the metallicity range of objects observed, as gauged by log(O/H)+12. 
This is in contrast to what is anticipated by recent model results. 
Post-AGB stellar models and the consequent PN abundance predictions published by \citet{karakas10} for low stellar metallicity progenitor stars [e.g. 0.004 (SMC), 0.008 (LMC), corresponding respectively to 8.00 and 8.30 on the horizontal scale) with birth masses below 4~M$_{\odot}$ suggest that C/O should be noticeably greater than in the case of the solar-like metallicities characteristic of our program objects. 
Similarly, above this mass threshold, where hot bottom burning (HBB) converts dredged up C into primary N, N/O should be greater at low metallicities than at levels closer to solar. 
HBB most likely occurred during the late evolutionary stages of progenitors of IC~2165, NGC~2440, NGC~5315 and PB6 since their N/O ratios are significantly higher than those of the remaining sample PNe. 
Note that the first three members of this group are {\it not} outstanding in terms of their C/O or He/H abundances. 
All of our objects were chosen so as to have near-solar metallicities, and the range of that parameter among them is rather small. 
Thus the lack of a metallicity trend among them in either C/O or N/O is not surprising. 
However, even with the extension of the sampled metallicity range provided especially by the LMC and SMC objects, the model-predicted trend remains elusive. 
This is perhaps due to the large amount of scatter observed among objects of similar metallicity (see below). 

Three of our eight objects (IC~2165, NGC~7662 and PB6) show C/O ratios which are greater than or equal to the solar value. 
In the cases of IC~2165 and PB6 the C/O ratio exceeds unity, suggesting that their progenitor AGB stars experienced significant third dredge-up during which fresh carbon produced by triple-alpha processing was brought up by convection into the stellar atmosphere from the He-burning shell located just above the C-O core of the remnant star.  

The left-hand panels in Fig.~\ref{fig:4plex} show the distribution of our program objects and the KH database objects in the log(C/O) and  log(N/O)-He/H planes. 
Note the suggestion of a positive correlation between log(N/O) and He/H. 
This has been noticed many times before; one such example is included in the extensive review by \citet{kaler85}. 
This positive behavior is predicted by stellar evolution models \citep[cf.][]{marigo01,karakas10} and is fundamentally associated with a positive correlation between N/O and stellar mass. 
In a followup paper to this one we explore more closely the relation between the progenitor masses and the abundance results of the present work.

\citet{peimbert83} referred to those PNe in which He/H$\ge$0.125 and log(N/O)$\ge$-0.30 as Type~I objects, while PNe below these boundaries were considered to be non-Type~Is. 
By these criteria then, NGC~2440 and PB6 are Type~I objects, while the other six are non-Type~Is. 
PB6's Type~I membership is confirmed by \citet{pena98}, who find He/H = 0.17 and log(N/O) = 0.14.

The PNe with supersolar N/O have several other interesting properties that distinguish them from those PNe with solar or subsolar N/O. 
The most dramatic are their morphologies. 
Members of the latter group are all elliptical in outline with very conspicuous central cavities surrounded by thin bright rims and smooth and sharply bounded shells. 
None of the PNe with supersolar N/O have similar morphologies. 
Either they are bipolar (NGC~2440), clumpy (PB6), or elliptical with central cavities but no thin rim (IC2165 and NGC5315). 
NGC5315 is also a serendipitously discovered diffuse x-ray source \citep{kastner08}. 
Thin rims are the result of relatively gentle expansions of the hot cavities within them, whereas fractured rims are the result of instabilities induced by rapid cavity expansion at high pressure \citep{toala14}. 
We note that as a rule, the nebular expansion velocities of PNe with [WC] nuclei are well above average \citep{gorny95}, again suggesting a high degree of momentum transfer from stellar winds during their evolutions.

The central stars of PNe with supersolar N/O also stand out from the others. 
NGC~5315 and PB6 have early-type WC or WO nuclei [see \citet{kaler91}, \citet{pena98} and  \citet{acker03}]. 
The central star of NGC~2440 is the hottest known \citep{hh90}. 
NGC~5315 and NGC~2440 have probably evolved from relatively massive progenitors whose mass loss rates in radiation-driven winds are or have been very strong. The central star of IC~2165, whose progenitor is probably not as massive as these, may simply be at a high-temperature stage in its evolution.

Finally, an obvious characteristic in each of the four panels in Fig.~\ref{fig:4plex} is the large amount of point-to-point scatter in C/O, N/O and He/H; scatter is present in all of the individual samples included in the figures. 
The sizes of the uncertainties (not indicated here to avoid visual confusion, but see Table~\ref{tab:ElemAbund}) are dwarfed by the sizable scatter exhibited in each panel. 
Therefore, the scatter is very likely real and indicative of the wide range of chemical inhomogeneity in the ISM from which these stars formed at various places and times within the galaxy and/or in the amount of atmospheric self-pollution that AGB stars experience during the final evolutionary stages of these stars.

\section{Conclusions}

The STIS long-slit data from this program are the first co-spatial spectra of extended Galactic PNe that span the UV and optical bands at sub-arcsecond spatial resolution. 
These new data enable a detailed and consistent analysis of abundance and physical properties of PNe using both UV  and optical emission lines from identically sampled volumes. 
Compared to prior abundance analyses of past decades that studied only the major optical emission lines, or that used satellite UV with ground-based data from different nebular regions, this work offers new insights on UV-optical emission variations and permits corresponding analysis of nebular diagnostics in C and N lines from their major ions. 
This initial study of these STIS data touches only the top of the science inherent in the data, which will be analyzed in more detail in future papers. 
These \textit{HST} data are now in the public domain and offer future investigators a new and, perhaps, historical insight into the spatial variations of UV-optical emission for modeling physical diagnostics across the extent of photoionized nebulae. 

We conclude that the central stars and the morphological outcomes of nebular evolution of PNe with super-solar N/O are not typical of those of PNe with solar N/O. 
Although the evidence is somewhat circumstantial in this paper, there is every reason to suppose that the central stars of PNe with super-solar N/O have evolved from some of the most massive stars that are able to form PNe. 
This is one of the reasons that we selected a PN sample with a wide range of N/O at constant O/H.  
It is therefore peculiar that PNe with super-solar N/O show no signs of C/O anomalies. 
Perhaps this can be explained by the conversion of C to N during HBB, thereby increasing the N while holding C to a level close to its original one in the progenitor star. 
On the other hand, the abundance trends in our graphs plotted against He/H indicate that selecting PN samples based on He enrichment will be a fruitful approach to understanding CNO production.

Finally, the large amounts of scatter in the N/O and C/O ratios at roughly constant metallicity is much larger than can be explained by observational uncertainties, (although we cannot rule out the possibility of systematic errors in earlier publications because of, e.g. mis-matched apertures between spectra from multiple wavelength regimes). 
In fact the size of the scatter in our plot of C/O or N/O versus O/H dwarfs that observed among objects such as \ion{H}{2} regions or main sequence stars, regardless of metallicity. 
We surmise that this situation may reflect the self-polluting nature of objects in the AGB stage of evolution or the chemical history of the Galactic ISM at the time and place when the stars formed \citep{matteucci03}

\acknowledgments
Support for Program number GO--12600 was provided by NASA through a grant from the Space Telescope Science Institute, which is operated by the Association of Universities for Research in Astronomy, Incorporated, under NASA contract NAS5--26555. 
B.B. received partial support from NSF grant AST--0808201. 


\clearpage


%
\begin{deluxetable}{crrlrc}
\tabletypesize{\footnotesize}
\tablecolumns{6}
\tablewidth{0pt}
\tablecaption {Slit Reference Positions and Extraction Regions \label{tab:Apertures}}
\tablehead {
\colhead{} & \multicolumn{2}{c}{Slit Reference Position} & \colhead {} & \colhead {} & \colhead {} \\*
\cline{2-3} \\*
\colhead{} & \colhead {RA} & \colhead {Dec} & \colhead {Offset from CS} & \colhead {Orient} & \colhead {Extract Bounds\tablenotemark{a}} \\*
\colhead {Nebula} & \colhead {(J2000)} & \colhead {(J2000)} & \colhead {(arcsec)} & \colhead {(deg)} & \colhead {(arcsec)}  
}
\startdata  
IC2165  &  6 21 42.775 & $-12$ 59 12.96 & 0     &    68.650  & $-6.12:+4.68$ \\
IC3568  & 12 33 07.340 &  +82  33 50.40 & 0.5 E &    29.684  & $-2.52:-0.72$, $+0.10:+1.62$ \\
NGC2440 & 07 41 54.875 & $-18$ 12 29.97 & Unknown\tablenotemark{b} & $-13.350$ & $-9.80:+1.26$ \\
NGC3242 & 10 24 46.107 & $-18$ 38 37.14 & 4.5 S &  $-77.350$ & $-7.20:+9.00$ \\
NGC5315 & 13 53 56.980 & $-66$ 30 50.60 & 0.4 N &  $-85.345$ & $-1.80:+0.36$, $+0.99:+3.60$ \\

NGC5882 & 15 16 49.938 & $-45$ 38 57.19 & 0.0   &  $-90.350$ & $-2.70:+2.52$ \\
NGC6537 & 18 05 13.129 & $-19$ 50 34.62 & 0.0   & $-136.350$ & $-1.80:+2.16$ \\
NGC6778 & 19 18 24.939 & $-01$ 35 47.41 & 0.0   &    89.651  & $-3.24:-0.36$, $+0.72:-3.60$ \\
NGC7662 & 23 25 53.600 &  +42  32 00.50 & 5.5 S &   101.649  & $-2.52:+5.40$ \\
PB6     & 10 13 15.989 & $-50$ 19 59.13 & 0.0   &    34.646  & $-6.48:-1.08$, $+0.00:+6.12$ \\
\enddata
\tablenotetext{a}{Extraction window along the slit, relative to reference position; multiple extractions were summed.} 
\tablenotetext{b}{Target acquisition failed: position uncertain and optical position differs from UV.} 
\end{deluxetable}

%
\begin{deluxetable}{ccccccr}
\tabletypesize{\scriptsize}
\tablecolumns{7}
\tablewidth{0pt}
\tablecaption {Observing Log for GO 126000 \label{tab:ObsLog}}
\tablehead {
\colhead{} & \colhead {} & \colhead {} & \colhead {Grating/} & \colhead {Dispersion} & \colhead {Slit Width\tablenotemark{a}} & \colhead {T$_{Exp}$} \\*
\colhead {Nebula} & \colhead {UT Date} & \colhead {Dataset} & 
\colhead {Cent. Wave} & \colhead {(\AA/pix)} & \colhead {(arcsec)} & \colhead {(s)}  
}
\startdata  
IC2165  & 2012-Apr-28 & OBRZ14010 & G140L/1425 & 0.60 & 0.2 & 2280 \\*
        &             & OBRZ14020 & G230L/2376 & 1.58 & 0.2 & 1946 \\*
        &             & OBRZ14030 & G230M/1884 & 0.09 & 0.5 & 346 \\*
        & 2012-Apr-30 & OBRZ13010 & G430L/4300 & 2.75 & 0.2 & 360 \\*
        &             & OBRZ13020 & G430M/4451 & 0.28 & 0.5 & 210 \\*
        &             & OBRZ13040 & G750L/7751 & 4.92 & 0.2 & 360 \\*
        &             & OBRZ13030 & G750M/6581 & 0.56 & 0.5 & 165 \\

IC3568  & 2012-Jul-15 & OBRZ02010 & G140L/1425 & 0.60 & 0.2 & 2624 \\
\enddata
\tablenotetext{a}{Slit width of 0.2\arcsec corresponds to 3.94 pix for the CCD, and 8.13 pix for the MAMA detectors; slit width of 0.5\arcsec corresponds to 9.85 pix for the CCD, and 20.33 pix for the MAMA. As described in the \textit{STIS Instrument Handbook} \citep{STIS_ihb}, for extended sources the spectral resolution is limited by the slit width for all settings.}
\end{deluxetable} 

\begin{landscape}
\begin{deluxetable}{rlr rr@{}l rr@{}l rr@{}l rr@{}l rr@{}l rr@{}l}
\tabletypesize{\scriptsize}
\setlength{\tabcolsep}{0.07in}
\tablecolumns{21}
\tablewidth{0in}
\tablecaption{Fluxes and Intensities I. \label{tab:Flux1}}
\tablehead{
\colhead{Wave} & \colhead{} & \colhead{} &
\multicolumn{3}{c}{IC2165} &
\multicolumn{3}{c}{IC3568} &
\multicolumn{3}{c}{NGC2440 UV} &
\multicolumn{3}{c}{NGC2440 Opt} &
\multicolumn{3}{c}{NGC3242} &
\multicolumn{3}{c}{NGC5315} \\
\cline{4-21}  \\
\colhead{(\AA)} & \colhead{ID\tnm{a}} & \colhead{$f$($\lambda$)} & 
\colhead{F($\lambda$)} &
\multicolumn{2}{c}{I($\lambda$)} &
\colhead{F($\lambda$)} &
\multicolumn{2}{c}{I($\lambda$)} &
\colhead{F($\lambda$)} &
\multicolumn{2}{c}{I($\lambda$)} &
\colhead{F($\lambda$)} &
\multicolumn{2}{c}{I($\lambda$)} &
\colhead{F($\lambda$)} &
\multicolumn{2}{c}{I($\lambda$)} &
\colhead{F($\lambda$)} &
\multicolumn{2}{c}{I($\lambda$)}
}
\startdata
1175 & C III             & 1.849  & 2.03         & 16.0    & $\pm$1.1         & 3.18\tnm{**}    & 7.05    & $\pm$1.11\tnm{**} & 1.30         & 10.1    & $\pm$0.6          & \nodata      & \nodata &                   & 27.4         & 30.7    & $\pm$0.8         & \nodata       & \nodata  & \\
1241 & N V               & 1.636  & 7.07         & 44.1    & $\pm$2.4         & 22.2\tnm{**}    & 45.0    & $\pm$2.8\tnm{**}  & 20.9         & 128.    & $\pm$2.           & \nodata      & \nodata &                   & 1.84         & 2.03    & $\pm$0.17        & 0.937\tnm{**} & 6.94    & $\pm$5.07\tnm{**} \\
1247 & C III             & 1.620  & \nodata      & \nodata &                  & \nodata         & \nodata &                   & \nodata      & \nodata &                   & \nodata      & \nodata &                   & 5.81         & 6.41    & $\pm$0.17        & 0.641\tnm{**} & 5.21    & $\pm$3.76\tnm{**} \\
1266 & Fe II             & 1.569  & \nodata      & \nodata &                  & \nodata         & \nodata &                   & \nodata      & \nodata &                   & \nodata      & \nodata &                   & 2.55         & 2.81    & $\pm$0.15        & \nodata       & \nodata &  \\
1305 & O I               & 1.478  & \nodata      & \nodata &                  & \nodata         & \nodata &                   & \nodata      & \nodata &                   & \nodata      & \nodata &                   & 4.38         & 4.79    & $\pm$0.18        & 0.353         & 2.39    & $\pm$0.22 \\
1324 & [Mg V]?           & 1.438  & \nodata      & \nodata &                  & \nodata         & \nodata &                   & 0.795        & 3.91    & $\pm$0.58         & \nodata      & \nodata &                   & \nodata      & \nodata &                  & \nodata       & \nodata &  \\
\enddata
\tablenotetext{**}{P Cyg profile; stellar line}
\tablenotetext{*}{P Cyg profile; stellar + nebular line contribution}
\tablenotetext{a}{Identifications ending in ``fl'' indicate fluorescence}
\tablenotetext{b}{Flux affected by possible artifact}
\tablenotetext{c}{Flux partitioned among multiple contributing emission lines: see text}
\tablenotetext{d}{{L}ogarithmic extinction at H$\beta$}
\tablenotetext{e}{Expected intrinsic H$\alpha$/H$\beta$ ratio at nebular temperature and density}
\tablenotetext{f}{Flux of H$\beta$ $\lambda4861$ within extraction aperture in ergs\ cm$^{-2}$ s$^{-1}$}
\end{deluxetable}

\begin{deluxetable}{rlc rr@{}l rr@{}l rr@{}l rr@{}l rr@{}l}
\tabletypesize{\scriptsize}
\tablecolumns{18}
\tablewidth{0in}
\tablecaption{Fluxes and Intensities II. \label{tab:Flux2}}
\tablehead{
\colhead{Wave} & \colhead{} & \colhead{} &
\multicolumn{3}{c}{NGC5882} &
\multicolumn{3}{c}{NGC6537} &
\multicolumn{3}{c}{NGC6778} &
\multicolumn{3}{c}{NGC7662} &
\multicolumn{3}{c}{PB6}\\
\cline{4-18}  \\
\colhead{(\AA)} & \colhead{ID\tnm{a}} & \colhead{$f$($\lambda$)} & 
\colhead{F($\lambda$)} &
\multicolumn{2}{c}{I($\lambda$)} &
\colhead{F($\lambda$)} &
\multicolumn{2}{c}{I($\lambda$)} &
\colhead{F($\lambda$)} &
\multicolumn{2}{c}{I($\lambda$)} &
\colhead{F($\lambda$)} &
\multicolumn{2}{c}{I($\lambda$)} &
\colhead{F($\lambda$)} &
\multicolumn{2}{c}{I($\lambda$)} 
}
\startdata
1175 & C III            &  1.849 & 0.508         & 2.46    & $\pm$0.56         & \nodata      & \nodata &                  & \nodata      & \nodata &                  & 12.6         & 17.1    & $\pm$1.2         & 2.81         & 9.72    & $\pm$13.19 \\
1241 & N V              &  1.636 & 2.32\tnm{**}  & 9.35    & $\pm$1.40\tnm{**} & \nodata      & \nodata &                  & \nodata      & \nodata &                  & 2.42         & 3.17    & $\pm$0.19        & 41.2         & 123.    & $\pm$136. \\
1247 & C III            &  1.620 & \nodata       & \nodata &                   & \nodata      & \nodata &                  & \nodata      & \nodata &                  & 0.297        & 0.388   & $\pm$0.050       & \nodata      & \nodata & \nodata \\
1266 & Fe II            &  1.569 & \nodata       & \nodata &                   & \nodata      & \nodata &                  & \nodata      & \nodata &                  & 1.04         & 1.35    & $\pm$0.11        & \nodata      & \nodata & \nodata \\
1305 & O I              &  1.478 & 0.212         & 0.747   & $\pm$0.278        & \nodata      & \nodata &                  & \nodata      & \nodata &                  & \nodata      & \nodata &                  & 130.         & 350.    & $\pm$356. \\
1336 & C II             &  1.415 & 0.618\tnm{**} & 2.06    & $\pm$0.23\tnm{**} & \nodata      & \nodata &                  & 1.07         & 11.8    & $\pm$18.1        & 2.25         & 2.85    & $\pm$0.18        & \nodata      & \nodata & \nodata \\
1344 & O IV + N II      &  1.400 & 0.253         & 0.835   & $\pm$0.248        & \nodata      & \nodata &                  & \nodata      & \nodata &                  & 0.663        & 0.835 & $\pm$0.125         & \nodata      & \nodata & \nodata \\
1370 & O V              &  1.354 & 0.564\tnm{**} & 1.79    & $\pm$0.18\tnm{**} & \nodata      & \nodata &                  & \nodata      & \nodata &                  & \nodata      & \nodata &                  & \nodata      & \nodata & \nodata \\
\enddata
\tablenotetext{**}{P Cyg profile; stellar line}
\tablenotetext{a}{Identifications ending in ``fl'' indicate fluorescence}
\tablenotetext{b}{Flux affected by possible artifact}
\tablenotetext{d}{Flux partitioned among multiple contributing emission lines: see text}
\tablenotetext{d}{{L}ogarithmic extinction at H$\beta$}
\tablenotetext{e}{Intrinsic H$\alpha$/H$\beta$ ratio at nebular temperature and density}
\tablenotetext{f}{Flux of H$\beta$ $\lambda4861$ within extraction aperture in ergs\ cm$^{-2}$ s$^{-1}$}
\end{deluxetable}
\end{landscape}

\begin{deluxetable}{cl}
\tabletypesize{\small}
\setlength{\tabcolsep}{0.07in}
\tablecolumns{2}
\tablewidth{0in}
\tablecaption{Unidentified Lines \label{tab:ULines}}
\tablehead{
\colhead{Wavelength} & \\
\colhead{(\AA)} & {Nebula} 
}
\startdata
1256		&	NGC~3242\\
1273		&	NGC~2440\\
1281		&	NGC~3242\\
1288		&	NGC~2440\\
1296		&	NGC~2440, NGC~3242\\       
1309		&	NGC~2440\\
1317	    	&	NGC~2440\\
1388		&	NGC~3242\\
1511		&	NGC~3242, NGC~7662\\
1591		&	NGC~3242\\
1981		&	NGC~3242\\
2280		&	NGC~3242\\
2932		&	NGC~2440\\
9346		&	NGC~6537\\
\enddata
\end{deluxetable}

\begin{landscape}
\begin{deluxetable}{cc r@{}l r@{}l r@{}l r@{}l r@{}l r@{}l r@{}l r@{}l r@{}l}
\tabletypesize{\scriptsize}
\setlength{\tabcolsep}{0.07in}
\tablecolumns{20}
\tablewidth{0in}
\tablecaption{Temperatures and Densities \label{tab:Diags}}
\tablehead{
\colhead{Parameter} &
\colhead{Diagnostic} &
\multicolumn{2}{c}{IC2165} &
\multicolumn{2}{c}{IC3568} &
\multicolumn{2}{c}{NGC2440} &
\multicolumn{2}{c}{NGC3242} &
\multicolumn{2}{c}{NGC5315} & 
\multicolumn{2}{c}{NGC5882} &
\multicolumn{2}{c}{NGC6537} &
\multicolumn{2}{c}{NGC6778} &
\multicolumn{2}{c}{NGC7662} 
}
\startdata
T$_e$ & [O III]  & 14230        & $\pm$162   & 10970        & $\pm$205  & 15350   & $\pm$311         & 11940        & $\pm$74   & 9561    & $\pm$125        &  9872        & $\pm$189         &   16680 & $\pm$513        & 10000\tnm{a} &           & 13170        & $\pm$226 \\
      & [N II]   & 10300\tnm{a} &            & 10300\tnm{a} &           & 10180   & $\pm$452         & 10300\tnm{a} &           & 12860   & $\pm$635        & 10300\tnm{a} &                  &   22560 & $\pm$2501       & 10300\tnm{a} &           & 10300\tnm{a} & \\
      & [O II]   & 21500        & $\pm$12060 & \nodata      &           & 11390   & $\pm$3109        & \nodata      &           & \nodata &                 & \nodata      &                  & \nodata &                 & \nodata      &           & \nodata      & \\
      & [S III]  & \nodata      &            & \nodata      &           & 13790   & $\pm$2045\tnm{c} & \nodata      &           & 9867    & $\pm$159\tnm{c} & 16870        & $\pm$2280\tnm{b} &   20570 & $\pm$839\tnm{c} & \nodata      &           & 13260        & $\pm$2628\tnm{b} \\[0.1in]
N$_e$ & [S II]   & 1606         & $\pm$1111  & \nodata      &           & 4580    & $\pm$2365        & \nodata      &           & 8819    & $\pm$2578       &   750        & $\pm$818         &   13240 & $\pm$5124       &     919      & $\pm$1331 & \nodata      &  \\
      & [Cl III] & \nodata      &            & \nodata      &           & \nodata &                  & \nodata      &           & 60130   & $\pm$93870      & \nodata      &                  & \nodata &                 & \nodata      &           & \nodata      & \\
      & [C III]  & 7987         & $\pm$2472  & 3620         & $\pm$6363 & 6362    & $\pm$1816        & 3837         & $\pm$1025 & 43620   & $\pm$11210      & 29490        & $\pm$28620       & \nodata &                 & \nodata      &           & 3288         & $\pm$976 \\
\enddata
\tablenotetext{a}{Default value.}
\tablenotetext{b}{Used 9069.}
\tablenotetext{c}{Used 9532.}
\end{deluxetable}

\begin{deluxetable}{lr@{}l r@{}l r@{}l r@{}l r@{}l r@{}l r@{}l}
\tabletypesize{\scriptsize}
\setlength{\tabcolsep}{0.07in}
\tablecolumns{16}
\tablewidth{0in}
\tablecaption{Ionic Abundances\tnm{a} \label{tab:IonicAbund}}
\tablehead{
\colhead{Ion} &
\multicolumn{2}{c}{IC2165} &
\multicolumn{2}{c}{IC3568} &
\multicolumn{2}{c}{NGC2440} &
\multicolumn{2}{c}{NGC3242} &
\multicolumn{2}{c}{NGC5315} &
\multicolumn{2}{c}{NGC5882} &
\multicolumn{2}{c}{NGC7662} 
}
\startdata
He$^{+}$(5876)   & 5.70    & $\pm$0.57(-2) & 0.114      & $\pm$0.012    & 5.35    & $\pm$0.46(-2) & 6.88    & $\pm$0.22(-2) & 0.132      & $\pm$0.002    & 0.103      & $\pm$0.004    & 6.72       & $\pm$0.63(-2) \\*
He$^{+2}(4686)$  & 4.85    & $\pm$0.10(-2) & 3.42       & $\pm$0.52(-3) & 7.39    & $\pm$0.06(-2) & 4.58    & $\pm$0.02(-2) & \nodata    &               & 6.92       & $\pm$0.22(-3) & 5.47       & $\pm$0.04(-2) \\
\\     
C$^{+}$(2325)    & 2.98    & $\pm$0.20(-6) & 7.55       & $\pm$1.46(-7) & 5.29    & $\pm$0.17(-6) & 6.51    & $\pm$0.16(-7) & 1.22       & $\pm$0.05(-6) & 3.41       & $\pm$0.40(-7) & 8.47       & $\pm$0.47(-7) \\
C$^{+2}$(1909)   & 1.72    & $\pm$0.13(-4) & 1.55       & $\pm$0.26(-4) & 1.42    & $\pm$0.16(-4) & 1.65    & $\pm$0.08(-4) & 2.22       & $\pm$0.27(-4) & 7.83       & $\pm$2.06(-5) & 1.28       & $\pm$0.14(-4) \\*
C$^{+2}$(1907)   & 1.72    & $\pm$0.13(-4) & 1.55       & $\pm$0.26(-4) & 1.42    & $\pm$0.16(-4) & 1.65    & $\pm$0.07(-4) & 2.21       & $\pm$0.27(-4) & 7.82       & $\pm$2.06(-5) & 1.28       & $\pm$0.14(-4) \\*
C$^{+2}$(4267)   & \nodata &               & \nodata    &               & \nodata &               & \nodata &               & 4.96       & $\pm$1.32(-4) & \nodata    &               & \nodata    & \\*
C$^{+2}$(adopt)  & 1.72    & $\pm$0.13(-4) & 1.55       & $\pm$0.26(-4) & 1.42    & $\pm$0.16(-4) & 1.65    & $\pm$0.08(-4) & 2.31       & $\pm$0.27(-4) & 7.83       & $\pm$2.06(-5) & 1.28       & $\pm$0.14(-4) \\
\enddata
\tablenotetext{a}{Abundances relative to H$^+$; $nn.n(-k) == nn.n \times 10^{-k}$}
\tablenotetext{b}{Derived using Te[\ion{S}{3}]}.
\end{deluxetable}

\begin{deluxetable}{lr@{}l r@{}l r@{}l r@{}l r@{}l r@{}l r@{}l ll}
\tabletypesize{\scriptsize}
\setlength{\tabcolsep}{0.07in}
\tablecolumns{20}
\tablewidth{0in}
\tablecaption{Total Elemental Abundances\tnm{a} \label{tab:ElemAbund}}
\tablehead{
\colhead{Parameter} &
\multicolumn{2}{c}{IC2165} &
\multicolumn{2}{c}{IC3568} &
\multicolumn{2}{c}{NGC2440} &
\multicolumn{2}{c}{NGC3242} &
\multicolumn{2}{c}{NGC5315} &
\multicolumn{2}{c}{NGC5882} &
\multicolumn{2}{c}{NGC7662} &
\colhead{Solar} &
\colhead{Orion}
}
\startdata
He/H      & 0.106   & $\pm$0.006    & 0.118   & $\pm$0.012    & 0.127   & $\pm$0.005    & 0.115   & $\pm$0.002    & 0.132 & $\pm$0.002    & 0.110   & $\pm$0.004    & 0.122 & $\pm$0.006    & 8.50(-2) & 9.70(-2) \\[0.1in]
C/H (ICF) & 3.56    & $\pm$0.33(-4) & 1.65    & $\pm$0.30(-4) & 4.61    & $\pm$0.73(-4) & 2.82    & $\pm$0.14(-4) & 2.43  & $\pm$0.28(-4) & 8.63    & $\pm$2.26(-5) & 2.38  & $\pm$0.28(-4) & 2.69(-4) & 3.31(-4) \\*
C/H (sum) & 3.17    & $\pm$0.18(-4) & 1.56    & $\pm$0.26(-4) & 1.95    & $\pm$0.17(-4) & 1.90    & $\pm$0.08(-4) & 2.34  & $\pm$0.27(-4) & 7.86    & $\pm$2.06(-5) & 2.36  & $\pm$0.20(-4) & 2.69(-4) & 3.31(-4) \\*
C/O (ICF) & 1.28    & $\pm$0.06     & 5.59    & $\pm$0.66(-1) & 1.14    & $\pm$0.07     & 7.07    & $\pm$0.19(-1) & 6.65  & $\pm$0.57(-1) & 1.95    & $\pm$0.46(-1) & 6.70  & $\pm$0.45(-1) & 5.50(-1) & 6.16(-1) \\*
C/O (sum) & 1.14    & $\pm$0.09     & 5.29    & $\pm$0.96(-1) & 0.48    & $\pm$0.07     & 4.77    & $\pm$0.23(-1) & 0.64  & $\pm$0.08     & 1.78    & $\pm$0.49(-1) & 6.70  & $\pm$0.45(-1) & 5.50(-1) & 6.16(-1) \\[0.1in]

N/H (ICF) & 8.01    & $\pm$0.61(-5) & 1.31    & $\pm$0.69(-5) & 6.59    & $\pm$0.55(-4) & 4.00    & $\pm$1.09(-5) & 3.86  & $\pm$0.30(-4) & 5.76    & $\pm$1.35(-5) & 4.34  & $\pm$1.18(-5) & 6.76(-5) & 5.37(-5) \\*
N/H (sum) & 8.57    & $\pm$0.54(-5) & $<$3.58(-5) &           & 4.13    & $\pm$0.26(-4) & 8.09    & $\pm$0.44(-5) & 1.49  & $\pm$0.32(-4) & $<$3.67(-5) &           & 4.95  & $\pm$0.46(-5) & 6.76(-5) & 5.37(-5) \\*
N/O (ICF) & 2.89    & $\pm$0.15(-1) & 4.44    & $\pm$2.54(-2) & 1.63    & $\pm$0.28     & 1.00    & $\pm$0.28(-1) & 1.06  & $\pm$0.08     & 1.30    & $\pm$0.29(-1) & 1.22  & $\pm$0.34(-1) & 1.38(-1) & 1.00(-1) \\
\enddata
\tablenotetext{a}{Abundances relative to H; $nn.n(-k) == nn.n \times 10^{-k}$}
\end{deluxetable}
\end{landscape}

\begin{deluxetable}{ccccl}
\tabletypesize{\scriptsize}
\setlength{\tabcolsep}{0.07in}
\tablecolumns{5}
\tablewidth{0in}
\tablecaption{Abundance Comparison\tnm{a} \label{tab:AbundComp}}
\tablehead{
\colhead{Nebula} & \colhead{C} & \colhead{N} & \colhead{O} & \colhead{Reference} 
}
\startdata
IC~2165 & 3.17(-4) & 8.57(-5) & 2.78(-4) & This paper \\*
        & 6.81(-4) & 1.48(-4) & 2.03(-4) & \citet{BRD13} \\*
        & 4.80(-4) & 7.30(-5) & 2.70(-4) &  \citet{PBS04} \\*
        &  \nodata & 1.33(-4) & 3.11(-4) &  \citet{KHM03} \\[0.1in]

IC~3568 & 1.56(-4) & 1.31(-5)\tnm{b} & 2.95(-4) & This paper \\
        & 1.31(-4) & 3.23(-5) & 2.45(-4) & \citet{L04a,L04b} \\
        &  \nodata & 1.80(-5) & 3.77(-4) & \citet{HKB04} \\[0.1in]

NGC~2440 & 1.95(-4) & 4.13(-4) & 4.04(-4) & This paper \\
\enddata
\tablenotetext{a}{Abundances relative to H; $nn.n(-k) == nn.n \times 10^{-k}$}
\tablenotetext{b}{ICF value; only N$^+$ available}
\end{deluxetable}


\begin{landscape}
\begin{figure}
\includegraphics[angle=270, scale=0.8]{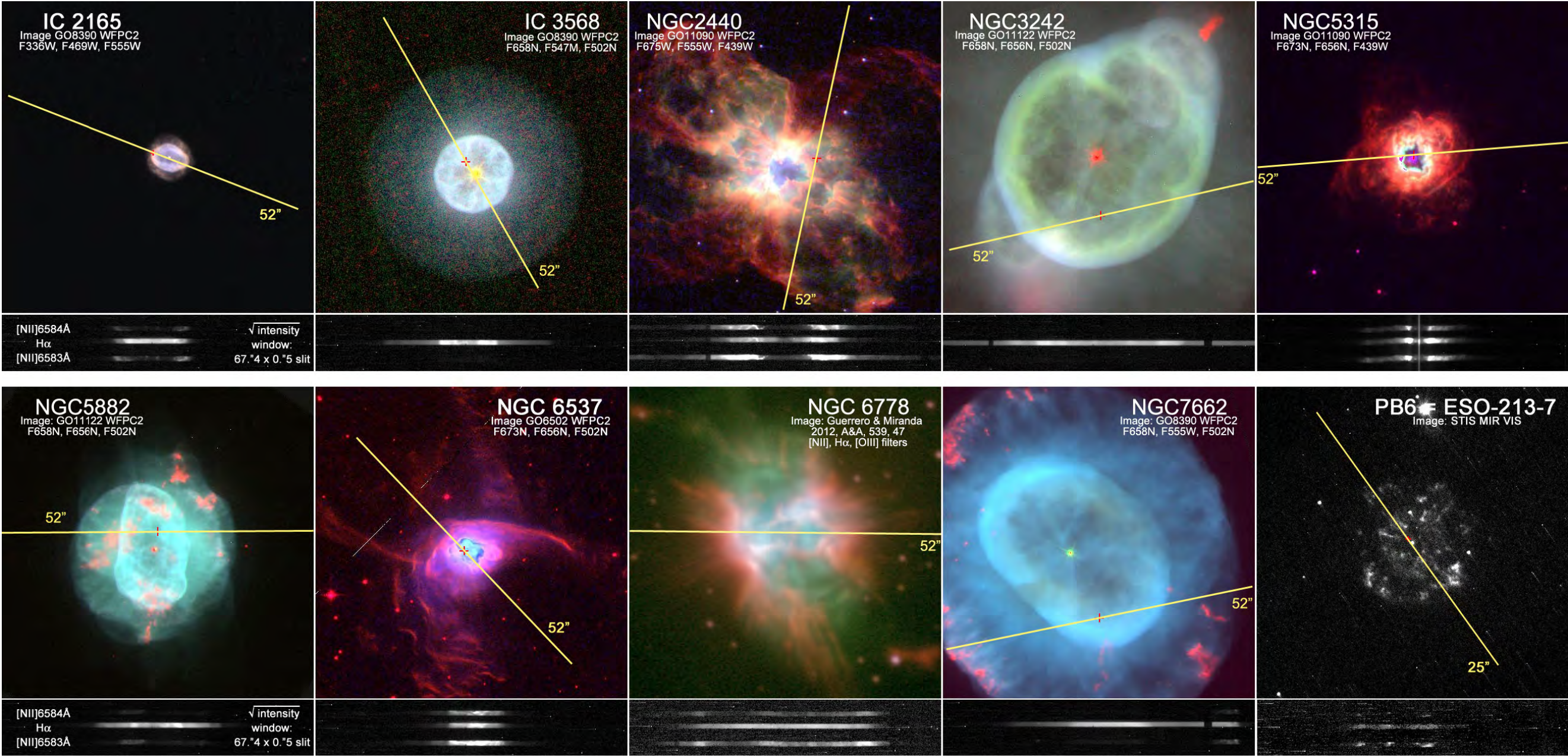}
\caption{
Slit positions and orientations (yellow lines) on multi-filter images of our target PNe. 
The image of PB6 is an exception: it is the pointing exposure for the STIS observations. 
Portions of the G750M STIS spectrum are shown below each image. 
They show the emission line intensity distributions of the H$\alpha$ and [\ion{N}{2}] lines that are characteristic of emission lines in our STIS data. 
These ``line profiles'' are convolved in the dispersion direction by thermal broadening and internal motions within the nebulae. 
The actual slit location of NGC~2440 and NGC~7662 may have been offset by up to 2\arcsec\ from the yellow line. 
\label{Montage}}
\end{figure}
\end{landscape}

\begin{figure}
\epsscale{1.0}
\plotone{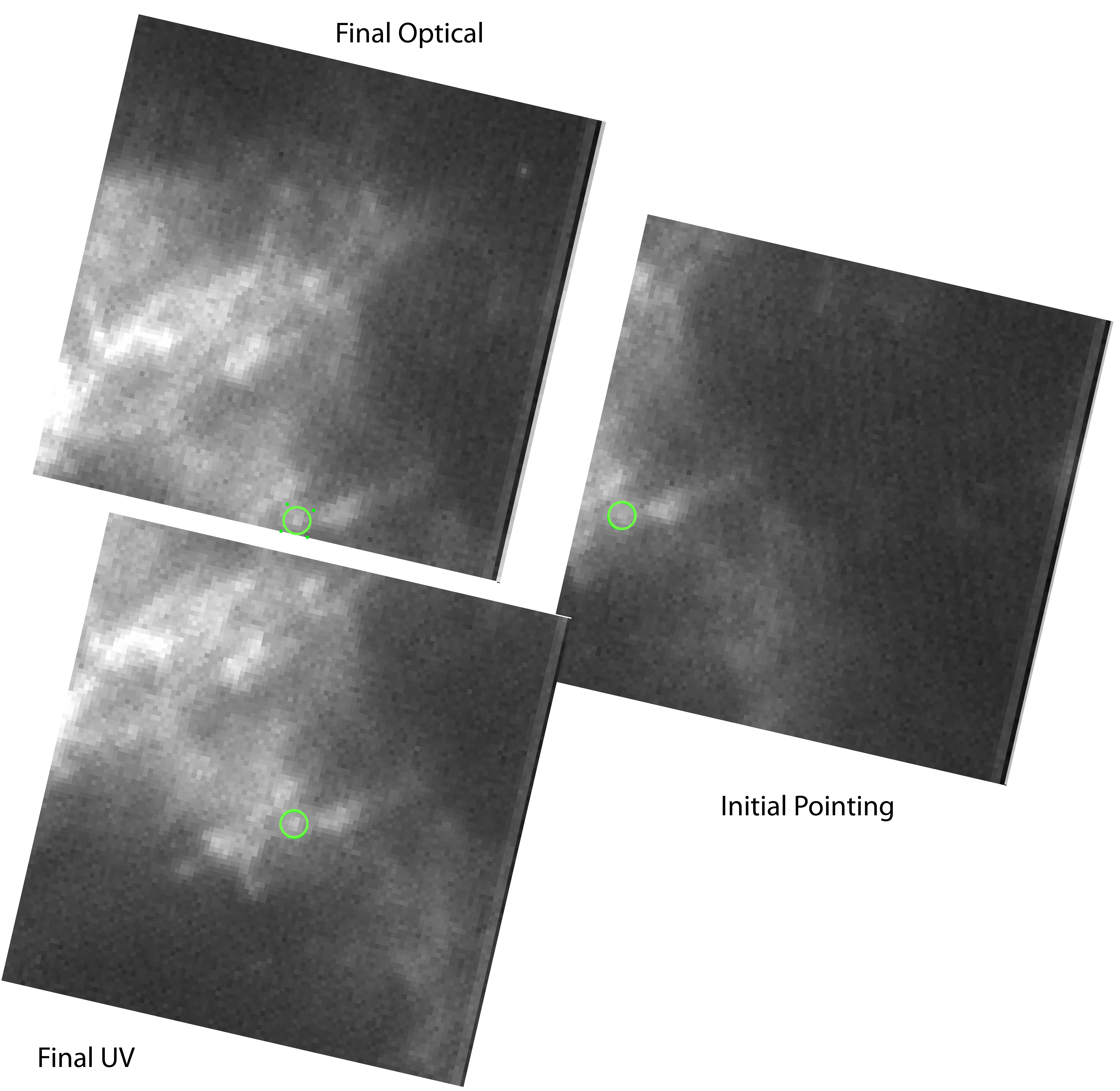}
\caption{
Acquisition images of NGC~2440 at the initial pointing, and the final positions for the UV and optical (northeast is in the upper-left). 
A green marker is superimposed on each image, indicating a nebular knot in common; the location of the knot indicates that the final location of the slit reference position in the optical-band lies almost directly south of the UV position by about 2.17\arcsec. 
\label{Acq_seq}}
\end{figure}

\begin{figure}
\includegraphics[angle=0, scale=0.70]{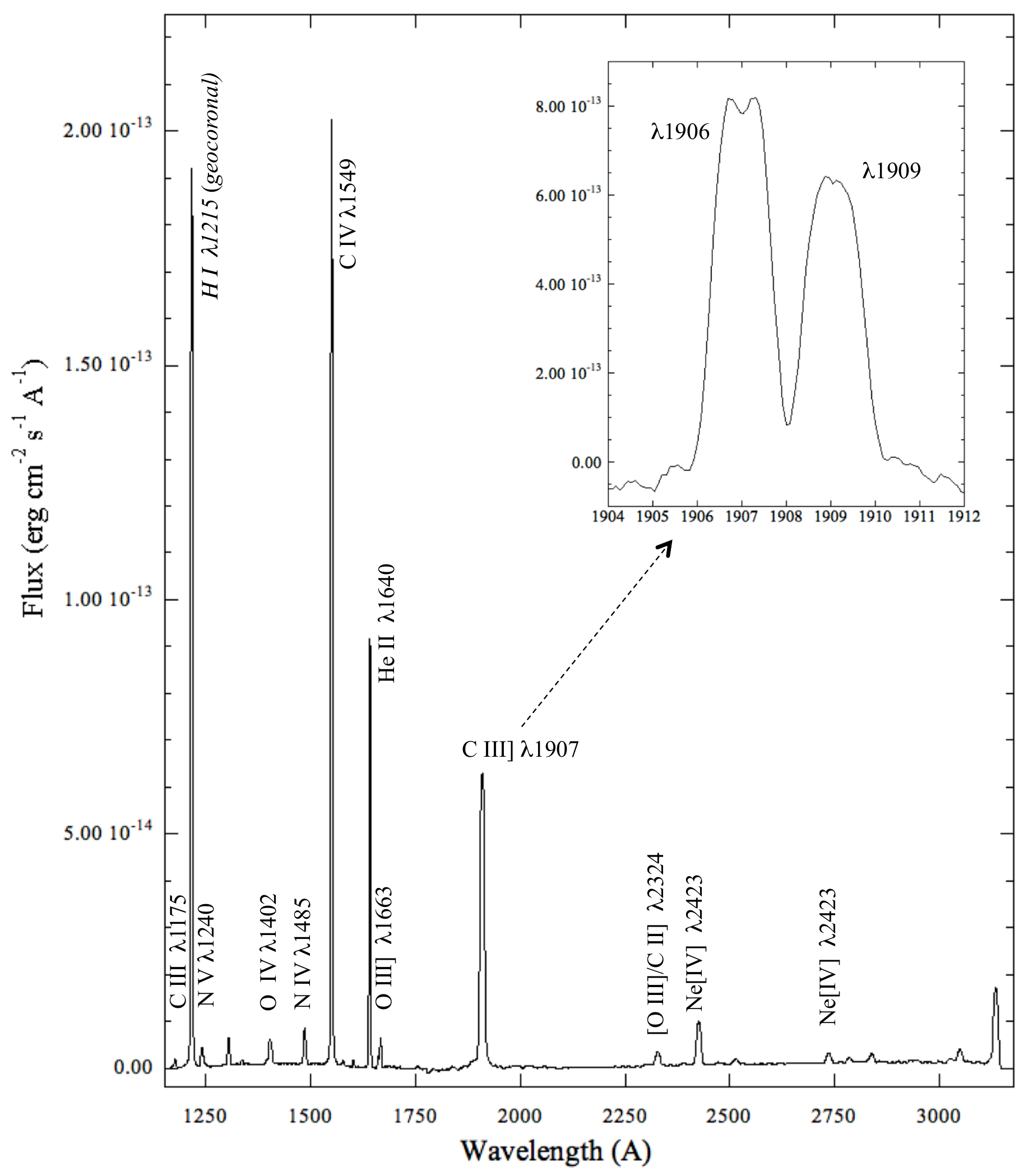}
\caption{
Spectrum of IC~2165 over the UV wavelength range. 
Numerous emission lines are identified. 
The inset shows an enlarged view of the closely-spaced lines of \ion{C}{3}] $\lambda\lambda$ 1906,1909. 
\label{spectrum1}}
\end{figure}

\begin{figure}
\includegraphics[angle=0, scale=0.70]{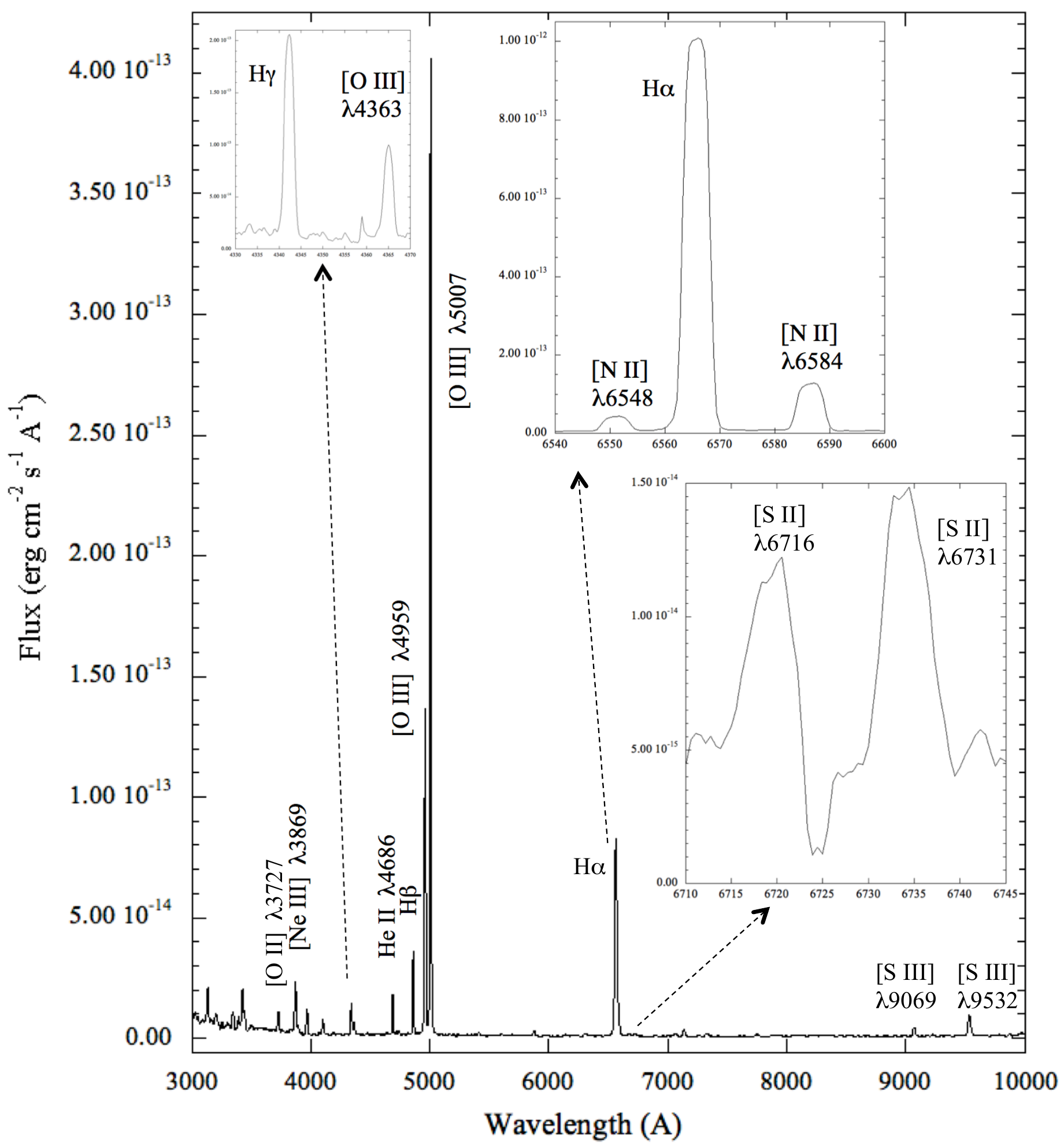}
\caption{
Spectrum of IC~2165 over the optical wavelength range. 
Numerous emission lines are identified. 
The insets show enlarged views of three line complexes: [\ion{S}{2}] $\lambda\lambda$6716,6731; [\ion{N}{2}] $\lambda\lambda$ 6548,6584 straddling H$\alpha$; and the region from H$\gamma$ to [\ion{O}{3}] $\lambda$4363. 
\label{spectrum2}}
\end{figure}

\begin{figure}
\epsscale{1.0}
\plotone{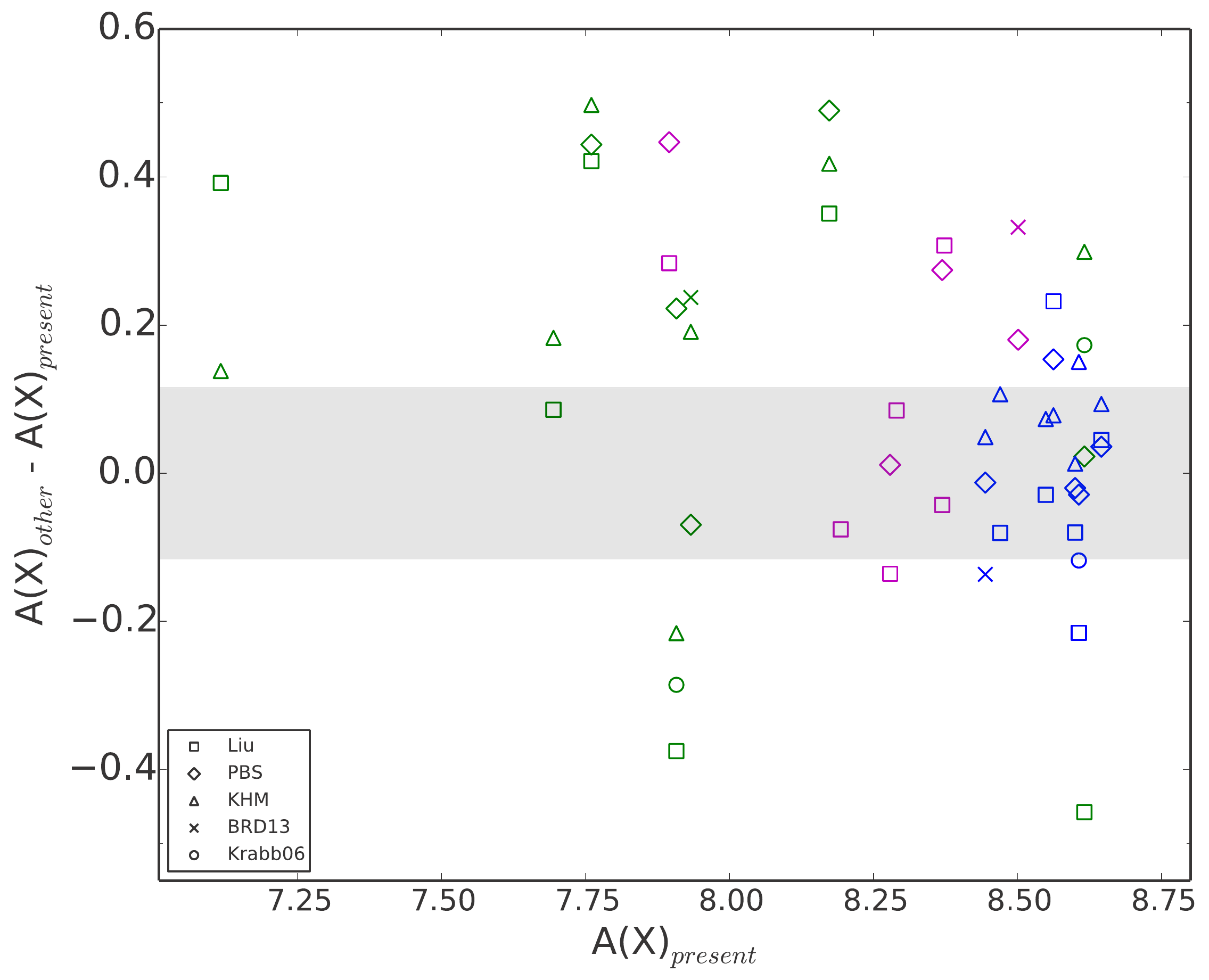}
 \caption{
 Difference between published log abundances and those of this study, as a function of our log abundance for the elements. Elements X are color-coded C: \textit{magenta}, N: \textit{green}, and O: \textit{blue}. 
 Published abundances are from 
Liu and collaborators (\textit{squares}): \citet{L04a,L04b, Tsamis03};
Pottasch and collaborators (\textit{diamonds}): \citet{P02, BS02, PBS08, PBS04, BRD13};
Kwitter and collaborators (\textit{triangles}): \citet{MHK02, KHM03, HKB04, Milingo10};
Dufour and collaborators (\textit{stars}): \citet{BRD13};
and Krabb (\textit{circles}): \citet{Krabbe06}.
Grey band shows agreement within 30\%. 
\label{AbundComp}}
\end{figure}

\begin{figure}
\plotone{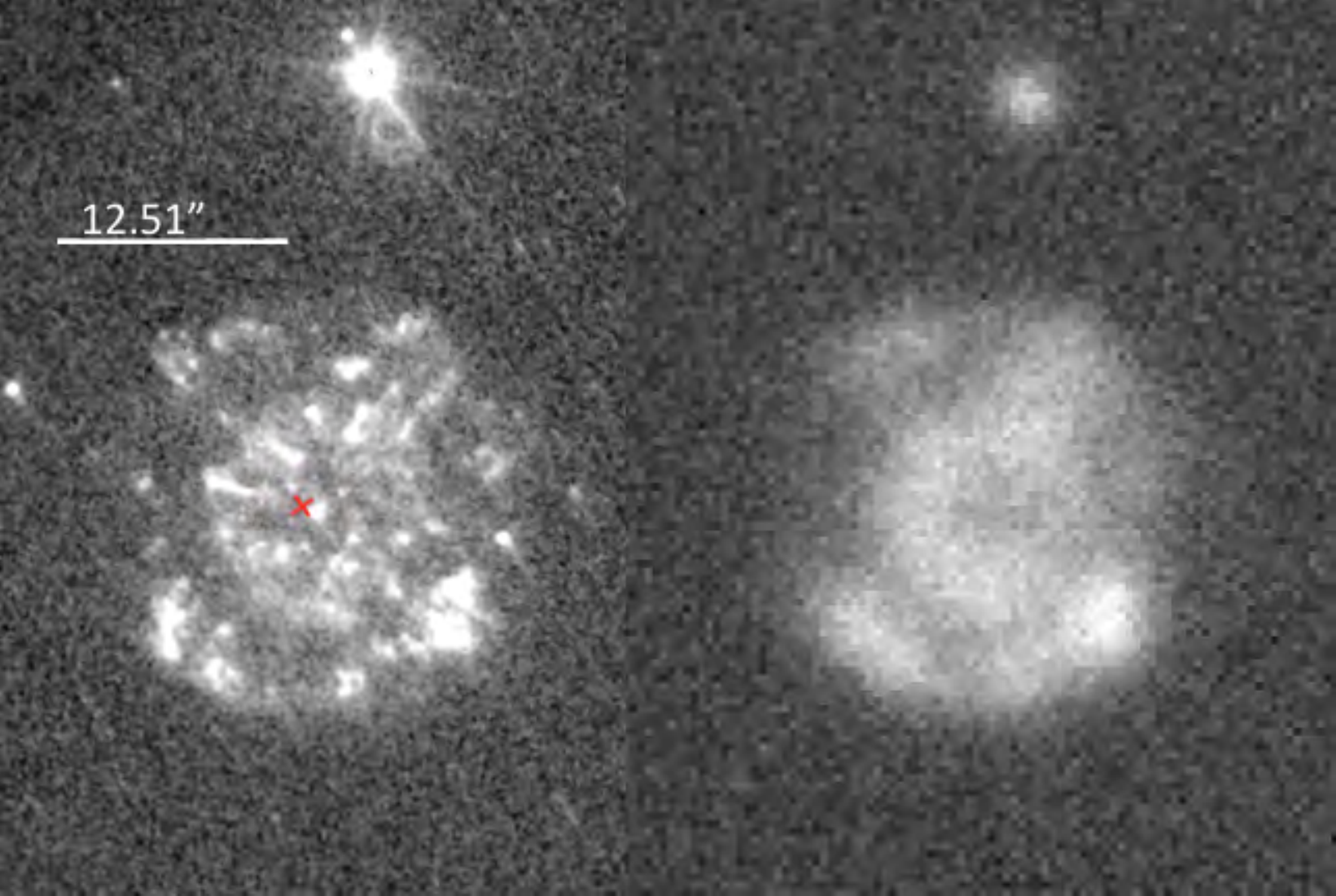}
 \caption{
Images of PB~6 obtained with STIS in white light through the F28X50LP aperture (\textit{left}), where the emission is dominated by H$\alpha$ and [\ion{N}{2}], compared to a ground-based image in [\ion{O}{3}] $\lambda5007$ obtained by one of us (R.L.M.C.) with the EFOSC2 camera on the 3.5m NTT ESO telescope under 1.5\arcsec\ seeing conditions. The images are oriented with North up and East to the left; the bar in the upper left provides the scale. 
Note that the apparent double-shell structure in the NTT image is resolved into a complex set of knots in the higher resolution STIS image. 
\label{fig:pb6}
}
\end{figure}

\begin{figure}
\includegraphics[angle=0, scale=0.60]{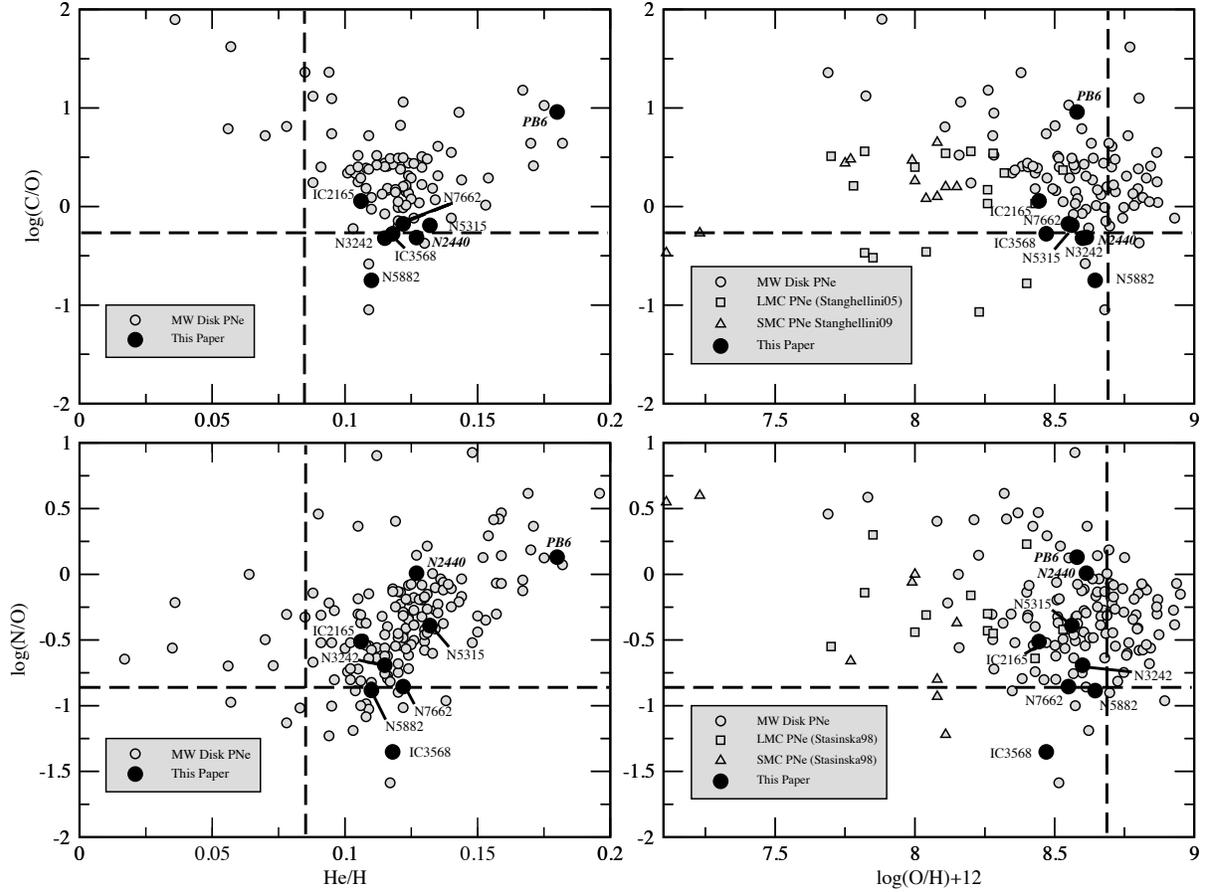}
 \caption{
 Left panels: log(C/O) and log(N/O) versus He/H, respectively. 
Program objects are represented by large filled circles, while data from the KH database are indicated with small gray circles. 
Right panels: Same as left panels but with log(O/H)+12 as the independent variable. 
Additional data for the LMC and SMC have also been included from papers cited in the legends. 
Bold dashed lines indicate solar values. 
Program objects are identified in each panel, where names of Type~I PNe appear in italics. 
Note that the abundances for PB~6 are taken from GRPP. 
\label{fig:4plex}
}
\end{figure}

\begin{thebibliography} {}

\bibitem[Acker \& Neiner(2003)]{acker03}
    Acker, A. \& Neiner, C.\ 2003, \aap, 403, 659

\bibitem[Asplund et al.(2009)]{A09}
    Asplund, M., Grevesse, N., Sauval, A.~J., \& Scott, P.\ 2009, \araa, 47, 481 

\bibitem[Bernard-Salas et al.(2002)]{BS02}
    Bernard-Salas, J., Pottasch, S.~R., Feibelman, W.~A., \& Wesselius, P.~R.\ 2002, \aap, 387, 301 

\bibitem[Bohigas et al.(2013)]{BRD13}
    Bohigas, J., Rodr{\'{\i}}guez, M., \& Dufour, R.~J.\ 2013, \rmxaa, 49, 227 

\bibitem[Bostroem \& Proffitt(2011)]{STIS_dhb}
    Bostroem, K., \& Proffitt, C.\ 2011, STIS Data Handbook (Version 6.0; Baltimore: STScI) 

\bibitem[Corradi et al. (2000)]{corradi00} Corradi, R.L.M., Goncalves, D.R., villaver, E., Mampaso, A., Perinotto, M., Schwarz, H.E., \& Zanon, C. 2000, \apj, 535, 3

\bibitem[Corradi et al.(2003)]{Corradi03}
    Corradi, R.~L.~M., Sch\"{o}nberner, D., \& Steffen, M., \& Perinotto, M.\ 2003, \mnras, 340, 417

\bibitem[Dufour(1991)]{dufour91}
    Dufour, R.~J., 1991, \pasp, 103, 857

\bibitem[Ely, et al.(2011)]{STIS_ihb}
    Ely, J., et al.\ 2011, STIS Instrument Handbook (Version 11.0; Baltimore: STScI) 

\bibitem[Esteban et al.(2004)]{esteban04}
    Esteban, C., Peimbert, M., Garc\'{i}a-Rojas, J., Ruiz, M.~T., Peimbert, A., \& Rodr\'{i}guez, M.\ 2004, \mnras, 355, 229

\bibitem[Fang et al.(2014)]{Fang_etal14}
     Fang, X., et al.\ 2014, \apj, 797, 100

\bibitem[Garc\'{i}a-Rojas, Pe\~{n}a \& Peimbert(2009)]{grpp09}
    Garc\'{i}a-Rojas, J., Pe\~{n}a, M., \& Peimbert A. 2009, \aap, 496, 139 (GRPP)

\bibitem[G\'{o}rny \& Stasi\'{n}ska(1995)]{gorny95}
    G\'{o}rny, S.~K., \& Stasi\'{n}ska, G.\ 1995, \aap, 303, 893

\bibitem[Heap \& Hintzen(1990)]{hh90}Heap, S.R., \& Hintzen, P.\ 1990, \apj, 353..200H

\bibitem[Henry et al.(1996)]{HKH96}
    Henry, R.~B.~C., Kwitter, K.~B., \& Howard, J.~W.\ 1996, \apj, 458, 215

\bibitem[Henry et al.(2000)]{HKB00}
    Henry, R.~B.~C., Kwitter, K.~B., \& Bates, J.~A.\ 2000, \aj, 531, 928 

\bibitem[Henry et al.(2004)]{HKB04}
    Henry, R.~B.~C., Kwitter, K.~B., \& Balick, B.\ 2004, \aj, 127, 2284 

\bibitem[Holland, et al.(2014)]{STIS_fluxCal}
    Holland, S.~T., Alessandra, A., Bostroem, A., Oliveira, C., \& Proffitt, C.\ 2014, Instrument Science Report STIS 2014-02 (Baltimore: STScI) 

\bibitem[Hyung(1994)]{Hyung94}
    Hyung, S.\ 1994, \apjs, 90, 119

\bibitem[Johnson et al.(2006)]{J06}
    Johnson, M.~D., Levitt, J.~S., Henry, R.~B.~C., \& Kwitter, K.~B.\ 2006, 
    IAU Symp. 234, ed. M.~J. Barlow \& Roberto M\'{e}ndez, (Cambridge) p. 439

\bibitem[Kaler(1985)]{kaler85}
    Kaler, J.~B.\ 1985, \araa, 23, 89

\bibitem[Kaler(1986)]{kaler86}
    Kaler, J.~B.\ 1986, \apj, 308, 322

\bibitem[Kaler et al.(1991)]{kaler91}
    Kaler, J.~B., Shaw, R.~A., Feibelman, W.~A., \& Imhoff, C.~L.\ 1991, \pasp, 103, 67

\bibitem[Karakas(2010)]{karakas10}
    Karakas, A.~I. 2010, \mnras, 403, 1413

\bibitem[Kastner et al.(2008)]{kastner08}
    Kastner, J.~H.; Montez, R., Jr., Balick, B., \& De Marco, O.\ 2008, \apj, 672, 957

\bibitem[Keller, Bianchi, \& Maciel(2014)]{keller14}
    Keller, G.~R.; Bianchi, L., \& Maciel, W.~J.\ 2014, \mnras, 442, 1379

\bibitem[Kingsburgh \& Barlow(1994)]{kb94}
    Kingsburgh, R.~L., \& Barlow, M.~J.\ 1994, \mnras, 271, 257

\bibitem[Koeppen \& Aller(1987)]{koeppen87}
    Koeppen, J., \& Aller, L.~H.\ 1987, in Exploring the Universe with the IUE Satellite, ASSL 129, 589 

\bibitem[Krabbe \& Copetti(2006)]{Krabbe06}
    Krabbe, A.~C., \& Copetti, M.~V.~F.\ 2006, \aap, 450, 159 

\bibitem[Kwitter \& Henry(1996)]{KH96}
    Kwitter, K.~B., \& Henry, R.~B.~C.\ 1996, \apj, 473, 304

\bibitem[Kwitter \& Henry(1998)]{KH98}
   Kwitter, K.~B., \& Henry, R.~B.~C.\ 1998, \apj, 493, 247

\bibitem[Kwitter \& Henry(2001)]{KH01}
    Kwitter, K.~B., \& Henry, R.~B.~C.\ 2001, \apj, 562, 804

\bibitem[Kwitter et al.(2003)]{KHM03}
    Kwitter, K.~B., Henry, R.~B.~C., \& Milingo, J.~B.\ 2003, \pasp, 115, 80 

\bibitem[Liu et al.(2004a)]{L04a}
    Liu, Y., Liu, X.-W., Luo, S.-G., \& Barlow, M.~J.\ 2004, \mnras, 353, 1231

\bibitem[Liu et al.(2004b)]{L04b}
    Liu, Y., Liu, X.-W., Barlow, M.~J., \& Luo, S.-G.\ 2004, \mnras, 353, 1251

\bibitem[Liu et al.(2006)]{L06}
    Liu, X.-W., Barlow, M. J., Zhang, Y., Bastin, R. J. \& Storey, P. J. 2006, \mnras, 368, 1959

\bibitem[Marigo (2001)]{marigo01}
    Marigo et al., 2001, \aap, 370, 194

\bibitem[Matteucci(2003)]{matteucci03}
    Matteucci, F.\ 2003, Ap\&SS, 284, 539

\bibitem[Milingo et al.(2002)]{MHK02}
    Milingo, J.~B., Henry, R.~B.~C., \& Kwitter, K.~B.\ 2002, \apjs, 138, 285 

\bibitem[Milingo et al.(2010)]{Milingo10}
    Milingo, J.~B., Kwitter, K.~B., Henry, R.~B.~C., \& Souza, S.~P.\ 2010, \apj, 711, 619

\bibitem[Peimbert \& Torres-Peimbert(1983)]{peimbert83}
    Peimbert. M., \& Torres-Peimbert, S.\ 1983, in Planetary Nebulae, IAU Symposium 103, (Dordrecht: Reidel), p.~233

\bibitem[Pe\~{n}a et al.(1998)]{pena98}
    Pe\~{n}a, M., Stasi\'{n}ska G. , Esteban, C., Koesterke, L., Medina, S., and Kingsburgh, R.\ 1998, \aap, 337, 866 

\bibitem[Perinotto(1991)]{perinotto91}
    Perinotto, M. 1991, \apjs, 76, 687

\bibitem[Pottasch et al.(2004)]{PBS04}
    Pottasch, S.~R., Bernard-Salas, J., Beintema, D.~A., \& Feibelman, W.~A.\ 2004, \aap, 423, 593 

\bibitem[Pottasch et al.(2002)]{P02}
    Pottasch, S.~R., Beintema, D.~A., Bernard-Salas, J., Koornneef, J., \& Feibelman, W.~A.\ 2002, \aap, 393, 285 

\bibitem[Pottasch \& Bernard-Salas(2008)]{PBS08}
    Pottasch, S.~R., \& Bernard-Salas, J.\ 2008, \aap, 490, 715 

\bibitem[Rola \& Stasi\'{n}ska(1994)]{rola94}
    Rola, C., \& Stasi\'{n}ska G. 1994, \aap, 282, 199

\bibitem[Savage \& Mathis(1979)]{savage79}
    Savage, B.~D., \& Mathis, J.~S. 1979, \araa, 17, 73

\bibitem[Seaton(1979)]{seaton79}
    Seaton, M.~J.\ 1979, \mnras, 187, 73P 

\bibitem[Shaw et al.(2010)]{Shaw10}
     Shaw, R.~A., et al.\ 2010, \apj, 717, 562

\bibitem[Stanghellini et al.(2005)]{stanghellini05}
    Stanghellini, L., Shaw, R.~A., \& Gilmore, D. 2005, \apj, 622, 294

\bibitem[Stanghellini et al.(2009)]{stanghellini09}
    Stanghellini, L., Lee, T-H, Shaw, R.~A., Balick, B., \& Villaver, E. 2009, \apj, 702, 733

\bibitem[Stasi\'{n}ska et al.(1998)]{stasinska98}
    Stasi\'{n}ska, G., Richer, M.~G., \& McCall, M.~L. 1998, \aap, 336, 667

\bibitem[Toal\'{a} \& Arthur(2014)]{toala14}
    Toal\'{a}, J.~A., \& Arthur, S.~J. 2014, MNRAS, 443, 3486

\bibitem[Tsamis et al.(2003)]{Tsamis03}
    Tsamis, Y.~G., Barlow, M.~J., Liu, X.-W., Danziger, I.~J., \& Storey, P.~J.\ 2003, \mnras, 345, 186 

\end{thebibliography}
\end{document}